\documentclass[useAMS, usenatbib, preprint, 12pt]{aastex}
\usepackage{cite, natbib}
\usepackage{float}
\usepackage{amsmath}
\usepackage{hyperref}
\usepackage{epsfig}
\usepackage{cases}
\usepackage[section]{placeins}
\usepackage{graphicx, subfigure}
\usepackage{color}
\usepackage{bm}

\AtBeginDocument{}

\newcommand{\ie}{{\it i.e.}}
\newcommand{\eg}{{\it e.g.}}

\newcommand{\kepler}{{\it Kepler}}
\newcommand{\Kepler}{{\it Kepler}}

\newcommand{\Ktwo}{{\it K2}}
\newcommand{\ktwo}{\Ktwo}

\newcommand{\tess}{{\it TESS}}

\newcommand{\lsst}{{\it LSST}}

\newcommand{\wfirst}{{\it WFIRST}}

\newcommand{\plato}{{\it PLATO}}
\newcommand{\Gaia}{{\it Gaia}}
\newcommand{\gaia}{{\it Gaia}}

\newcommand{\teff}{$T_{\mathrm{eff}}$}

\newcommand{\feh}{[Fe/H]}
\newcommand{\prot}{$P_{\mathrm{rot}}$}
\newcommand{\pmega}{$\bar{\pi}$}

\newcommand{\logg}{log(g)}
\newcommand{\dnu}{$\Delta \nu$}
\newcommand{\numax}{$\nu_{\mathrm{max}}$}

\newcommand{\amnh}{1}
\newcommand{\cca}{2}
\newcommand{\florida}{3}
\newcommand{\hawaii}{4}
\newcommand{\columbia}{5}
\newcommand{\riverside}{6}
\newcommand{\cuny}{7}
\newcommand{\nyu}{8}
\newcommand{\cds}{9}
\newcommand{\mpia}{10}
\newcommand{\yale}{11}

\newcommand{\sd}{{\tt stardate}}

\newcommand{\gcolor}{$G_{BP} - G_{RP}$}

\begin{document}

\title{Towards precise stellar ages: combining isochrone fitting with
empirical gyrochronology}

\author{%
    Ruth Angus\altaffilmark{\amnh, }\altaffilmark{\cca, },
    Timothy D. Morton\altaffilmark{\florida, }\altaffilmark{\cca, },
    Daniel Foreman-Mackey\altaffilmark{\cca},
    Jennifer van Saders\altaffilmark{\hawaii},
    Jason Curtis\altaffilmark{\columbia},
    Stephen R. Kane\altaffilmark{\riverside},
    Megan Bedell\altaffilmark{\cca},
    Rocio Kiman\altaffilmark{\cuny, }\altaffilmark{\amnh},
    David W. Hogg\altaffilmark{\nyu, }\altaffilmark{\cca,
}\altaffilmark{\cds, }\altaffilmark{\mpia},
    John Brewer\altaffilmark{\yale}
}

\altaffiltext{\amnh}{American Museum of Natural History, Central Park West,
Manhattan, NY, USA}
\altaffiltext{\cca}{Center for Computational Astrophysics, Flatiron Institute,
162 5th Avenue, Manhattan, NY, USA}
\altaffiltext{\florida}{Department of Astronomy, University of Florida,
Gainesville, FL, USA}
\altaffiltext{\hawaii}{Institute for Astronomy, University of Hawai'i at
M\={a}noa, Honolulu, HI, USA}
\altaffiltext{\columbia}{Department of Astronomy, Columbia
University, Manhattan, NY, USA}
\altaffiltext{\riverside}{
Department of Earth and Planetary Sciences, University of California,
Riverside, CA 92521, USA}
\altaffiltext{\cuny}{Department of physics, CUNY Graduate Center, City
University of New York, Manhattan, NY, USA}
\altaffiltext{\nyu}{Center for Cosmology and Particle Physics, New York
University, Manhattan, NY, USA}
\altaffiltext{\cds}{Center Data Science, New York
University, Manhattan, NY, USA}
\altaffiltext{\mpia}{Max-Planck Institute for Astronomy,
K{\"o}nigstuhl, Heidelberg, Germany}
\altaffiltext{\yale}{Yale center for astronomy and astrophysics, Yale
University, New Haven, CT, USA}

\begin{abstract}
We present a new age-dating technique that combines gyrochronology with
isochrone fitting to infer ages for FGKM main-sequence and subgiant field
stars.
Gyrochronology and isochrone fitting are each capable of providing relatively
    precise ages for field stars in certain areas of the Hertzsprung-Russell
    diagram: gyrochronology works optimally for cool main-sequence stars, and
    isochrone fitting can provide precise ages for stars near the
    main-sequence turnoff.
Combined, these two age-dating techniques can provide precise and accurate
ages for a broader range of stellar masses and evolutionary stages than either
method used in isolation.
We demonstrate that the position of a star on the Hertzsprung-Russell or
color-magnitude diagram can be combined with its rotation period to infer a
precise age via both isochrone fitting and gyrochronology simultaneously.
We show that incorporating rotation periods with 5\% uncertainties into
    stellar evolution models improves age precision for FGK stars on the main
    sequence, and can, on average, provide age estimates up to three times
    more precise than isochrone fitting alone.
In addition, we provide a new gyrochronology relation, calibrated to the
Praesepe cluster and the Sun, that includes a variance model to capture the
rotational behavior of stars whose rotation periods do not lengthen with the
square-root of time, and parts of the Hertzsprung-Russell diagram where
gyrochronology has not been calibrated.
This publication is accompanied by an open source {\it Python} package, \sd,
for inferring the ages of main-sequence and subgiant FGKM stars from rotation
periods, spectroscopic parameters and/or apparent magnitudes and parallaxes.
\end{abstract}

\section{Introduction}
\label{section:intro}

Age is the most difficult stellar property to measure, and the difficulty of
age-dating is particularly acute for low mass (GKM) stars on the main sequence
(MS).
Using conventional dating methods, uncertainties on the ages of these stars
can be as large as the age of the Universe.
GKM dwarfs are difficult to age-date because most of their physical and
observable properties do not change rapidly.
This is represented in the spacing of isochrones on a Hertzsprung-Russell
diagram (HRD) or color-magnitude diagram (CMD).
On the MS, isochrones are tightly spaced and, even with very precise
measurements of effective temperature and luminosity, the position of a MS
star on the HRD may be consistent with range of isochrones spanning several
billion years \citep[see][for a review of stellar ages]{soderblom2010}.
At the main-sequence turnoff however, isochrones are spread further apart, so
that sufficiently precisely measured temperatures and luminosities can yield
ages that are extremely precise (with minimum statistical uncertainties on the
order of 5-10\%) \citep[\eg][]{pont2004}.
The classical method for measuring stellar ages is isochrone placement, or
isochrone fitting, where surface gravity changes resulting from fusion in the
core (usually observed via luminosity, $L$, and effective temperature, \teff,
or absolute magnitude and colour) are compared with a set of models that trace
stellar evolution across the HRD, or CMD.
CMD/HRD position has been thoroughly mapped with physical models, and can be
used to calculate relatively accurate (but not necessarily precise) ages,
barring some systematic variations between different models,
\citep[\eg][]{yi2001, dotter2008, dotter2016}.
On the MS itself, there is little differentiation between stars of different
ages in the $L$ and \teff\ plane, so ages tend to be very imprecise.
The method of inferring a star's age from its rotation period, called
`gyrochronology', is much better suited for measuring ages on the MS because
MS stars spin down relatively rapidly.

Magnetic braking in MS stars was first observed by \citet{skumanich1972} who,
studying young clusters and the Sun, found that the rotation periods of
Solar-type stars decay with the square-root of time.
It has since been established that the rotation period of a MS star depends,
to first order, only on its age and effective temperature or color
\citep[\eg][]{barnes2003}.
The convenient characteristic of stars that allows their ages to be inferred
from their {\it current} rotation periods and independently of their
primordial ones, comes from the steep dependence of spin-down rate on rotation
period \citep{kawaler1989}.
Stars spinning with high angular velocity will experience a much greater
angular momentum loss rate than slowly spinning stars and for this reason, no
matter the initial rotation period, Solar type stars will have the same
rotation period after around the age of the Hyades, 500-700 million years
\citep{irwin2009, gallet2015}.
After this time, the age of a star can be inferred, to first order, from its
dust-corrected color (\eg\ B-V or \gcolor) and {\it current} rotation period
alone \citep[See][for an analysis of how initial conditions effect gyrochronal
ages]{epstein2014}.

The relation between age, rotation period and mass has been studied in detail,
and several different models have been developed to capture the rotational
evolution of Sun-like stars.
Some of these models are theoretical and based on physical processes; modeling
angular momentum loss as a function of stellar properties as well as the
properties of the magnetic field and stellar wind \citep[\eg][]{kawaler1988,
kawaler1989, pinsonneault1989, vansaders2013, matt2015, vansaders2016}.
Other models are empirical and capture the behavior of stars from a purely
observational standpoint, using simple functional forms that can reproduce the
data \citep[\eg][]{barnes2003, barnes2007, mamajek2008, angus2015}.
Both types of model, theoretical and empirical, must be calibrated using
observations.
Old calibrators are especially important because new evidence suggests that
rotational evolution goes through a transition at old age or, more
specifically, at a large Rossby number, $Ro$ (the ratio of rotation period to
the convective overturn timescale).
For example, stars shown to be old from \kepler\ asteroseismic data rotate
more rapidly than expected given their age \citep{angus2015, vansaders2016}.
A new physically motivated gyrochronology model, capable of reproducing these
data, was recently introduced \citep{vansaders2016}.
It relaxes magnetic braking at a critical Rossby number of around 2,
approximately the Solar value.
This model predicts that, after stellar rotation periods lengthen enough to
move stars across this $Ro$ threshold, stars conserve angular momentum and
maintain a nearly constant rotation period from then until they evolve off the
MS.

The gyrochronology models that capture post $Ro$-threshold, rotational
evolution \citep{vansaders2016} are the current state-of-the-art in rotation
dating.
These models can be computed over a grid of stellar parameters, and
interpolated-over to predict the age of a star.
The process of measuring the age of a field star with these models is similar
to inferring an age using any set of isochrones, with the difference being
that rotation period is an additional observable dimension.
Ages calculated using these models are therefore likely to be much more
precise than using rotation-free isochrones since rotation period provides an
additional anchor-point for the age of a star.
We present here a complementary method that combines isochrones with an {\it
empirical} gyrochronology model.
The methodology is related to the \citet{vansaders2016} model in that both use
a combination of rotation periods and other observable properties that track
stellar evolution on the HRD in concert.
The main difference is that the gyrochronology model used here is an entirely
empirically calibrated one, as opposed to a physically derived one.

One major advantage of using a physically motivated gyrochronology model is
the ability to rely on physics to interpolate or extrapolate over parts of
parameter space with sparse data coverage.
However, rotational spin-down is a complex process that is not yet fully
understood and currently no physical model can accurately reproduce all the
data available.
For this reason, even physically motivated gyrochronology models cannot always
be used to reliably extrapolate into unexplored parameter space.
Physical models, when calibrated to data, can provide insight into the physics
of stars, however, if accurate and precise {\it prediction} of stellar
properties is desired, empirical models can have advantages over physical
ones.
For example, the data may reveal complex trends that cannot be reproduced with
our current understanding of the physical processes involved, but may be
captured by more flexible data-driven models.
In addition, it is relatively straightforward to build an element of
stochasticity into empirical models, \ie\ to allow for and incorporate
outliers or noisy trends.
This is particularly important for stellar spin down because rotation periods
can be affected by additional confounding variables which are not always
observed (having a binary or planetary companion, for example).
A further advantage of empirical models is that inference is more tractable:
it can be extremely fast to fit them to data.

In this work we calibrated a new empirical gyrochronology relation, fit to the
Praesepe open cluster and the Sun, in \gaia\ \gcolor\ color.
\Gaia\ $G$, $G_{BP}$ and $G_{RP}$ apparent magnitudes are now the most
abundant photometric measurements, available for more than a billion stars
\citep{brown2018}.
In fact, an important point of context for this work is the new
availability of data relevant to gyrochronology and stellar ages.
\Gaia\ now provides broad-band photometry and parallaxes for over a billion
stars, and \kepler, \ktwo\ and \tess\ are providing rotation periods for
hundreds of thousands of stars.
Gyrochronology is becoming one of the most readily available age-dating
methods, so continuing to improve gyrochronology relations and methods is
important.
Like any other gyrochronology relation, this new Praesepe-based model does not
perfectly reproduce all the observed data and some simple modifications could
make significant improvements, for example, by including a mixture model to
account for outliers and binaries, and by removing the period-age,
period-color separability to account for different period-color shapes seen in
clusters of different ages.
We leave these improvements for a future project and, for now, test this
new Praesepe-based model, which is built into an open source {\it Python}
package for gyrochronology called \sd.
\sd\ provides the framework for simultaneous gyrochronology and isochrone
fitting, and, because \sd\ is modular, it would be straightforward to update
the gyrochronology relation in the future.

This paper is laid out as follows.
In section \ref{section:method} we describe our new age-dating model and its
implementation, in section \ref{section:results} we test this model on
simulated stars and cluster stars, and in section \ref{section:conclusion} we
discuss the implications of these tests and future pathways for development.
Throughout this paper we use the term {\it `observables'} to refer to the
following observed properties of a star: \teff, \logg, observed bulk
metallicity ($[\mathrm{Fe/H}]$), parallax ($\bar{\pi}$), photometric
colors in different passbands (${\bf m_x} = [m_J, m_H, m_K, m_B, m_V, m_G,
m_{GBP}, m_{GRP}...]$, etc) and rotation period ($P_{\mathrm{rot}}$).
The term `parameters' refers to the physical properties of that star: age
($t$), equivalent evolutionary phase (EEP), model metallicity
($\mathrm{M/H}$), distance ($D$) and V-band extinction ($A_V$).
These are the properties that generate the observables.

\section{Method}
\label{section:method}
\subsection{A new empirical gyrochronology relation}

\begin{figure}
  \caption{
    The rotation periods of Praesepe members \citep{douglas2016},
    vs. their \Gaia\ colors (\gcolor) with a broken power law model, fit to
    these data.
    The dashed line and shaded region shows the mean and variance model
    described in equation \ref{eqn:gyro} at 650 Myrs (lower model) and 4.56
    Gyrs (upper model).
    The solid blue line shows the \citep{angus2015} gyrochronology relation at
    650 Myrs (lower model) and 4.56 Gyrs (upper model).
    Shaded regions show the 1$\sigma$ range of the rotation period
    model (equation \ref{eqn:gyro}).
    The rotation periods of stars bluer than around 0.56 dex and redder than
    around 2.7 dex in \gcolor\ are modeled as a broad log-normal distribution
    with a standard deviation of 0.5 dex, added to observational
    uncertainties.
    This figure was generated in a Jupyter notebook available at
    \url{https://github.com/RuthAngus/stardate/blob/master/paper/code/Fitting_Praesepe.ipynb}
}
  \centering
    \includegraphics[width=1.1\textwidth]{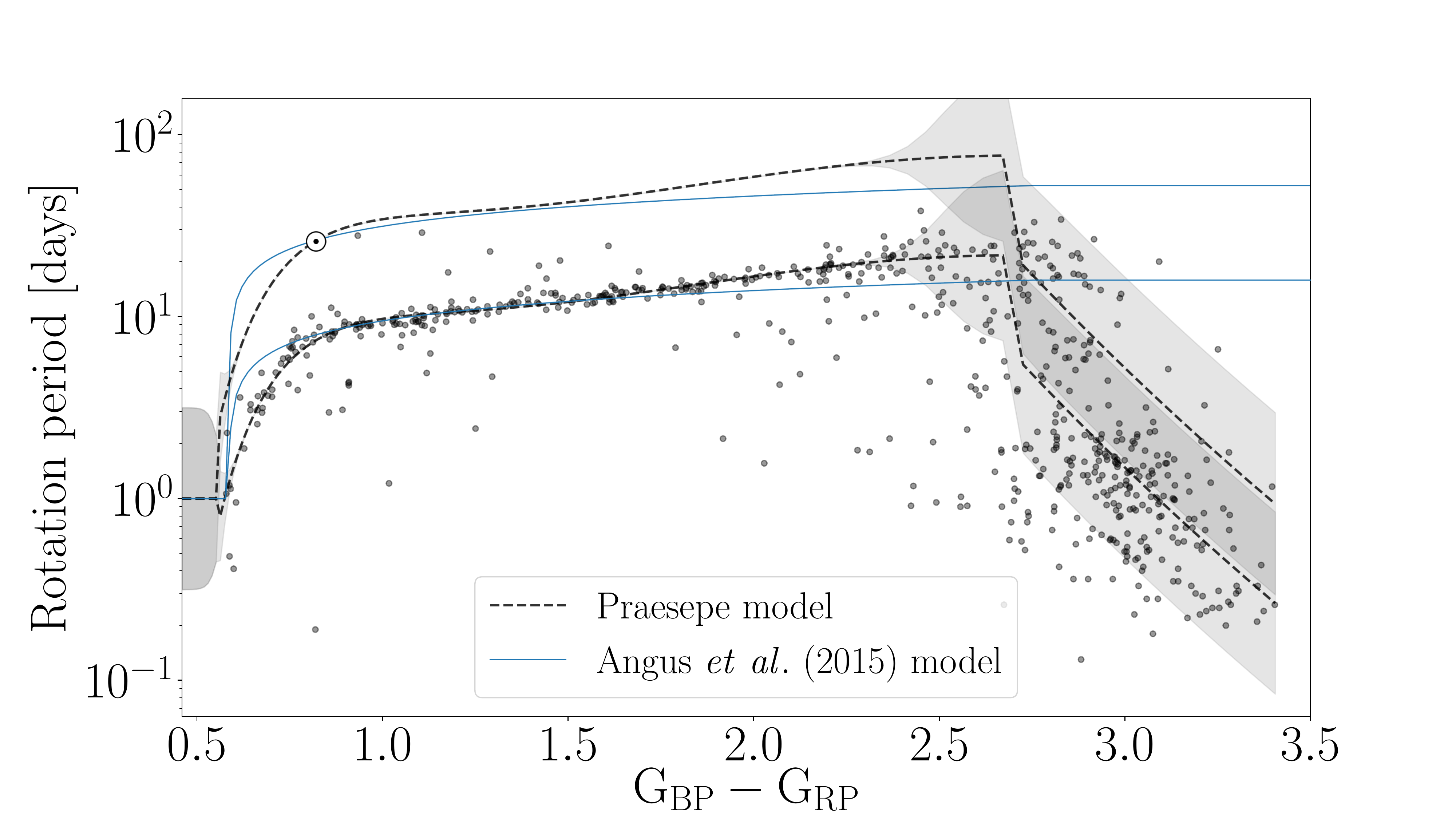}
\label{fig:praesepe}
\end{figure}

We fit a broken power law to the 650 Myr Praesepe cluster in order to
calibrate a gyrochronology relation that takes advantage of new data available
from the \ktwo\ and \gaia\ surveys.
This relation captures the detailed shape of Praesepe's rotation period-color
relation, and is calibrated to \gaia\ \gcolor\ color.
Praesepe is the ideal calibration cluster because it has the largest number of
members with precisely measured rotation periods, over a large range of colors
\citep[spanning spectral types A0 through M6,][]{rebull2017}, of any open
cluster.
It is relatively tightly clustered on the sky and many of its members were
targeted in a single \ktwo\ campaign, from which rotation periods have been
measured via light curve frequency analysis \citep{douglas2017, rebull2017}.
We compiled rotation periods, \Gaia\ photometry and \gaia\ parallaxes for
members of the 650 Myr Praesepe cluster \citep{fossati2008} by identifying
Praesepe members with measured rotation periods from \citet{douglas2017} in
the \ktwo-\gaia\ crossmatch catalog provided at
\url{https://gaia-kepler.fun/}.
This catalog cross-matched the EPIC catalog \citep{huber2016} with the
Gaia DR2 catalog \citep{brown2018}, using a 1'' search radius.
The result was a sample of 757 stars with rotation periods, parallaxes and
\gaia\ $G$, $G_{BP}$ and $G_{RP}$-band photometry, shown in figure
\ref{fig:praesepe}\footnote{This analysis was performed in a Jupyter notebook
available here: \url{
    https://github.com/RuthAngus/stardate/blob/master/paper/code/Praesepe.ipynb}}.
Although this new model does not perfectly describe the rotation period of
every star, it provides a better representation of rotational evolution than
previous models such as the \citet{angus2015} model (shown as the blue solid
line in figure \ref{fig:praesepe}).
In order to fit a relation to Praesepe, we removed rotational outliers bluer
than \gcolor\ = 2.7 via sigma-clipping and fit a 5th-order polynomial to the
remaining FGK and early M stars.
    We found that a 5th order polynomial provided a substantially
better fit than lower-order polynomials, which were not able to capture the
sharp `elbow’ in the rotation period-color relation. Additional orders
provided either a worse
fit, tending towards extreme values at the boundaries, or diminishing returns
in goodness-of-fit.
We also fit a straight line to the late M dwarfs (\gcolor\ $>$ 2.7), to
capture the mass-dependent initial rotation periods of low mass stars
\citep{somers2017}.
We fit a separable straight line function to the period-age relation using
the ages of Praesepe and the Sun\footnote{The fitting process was
performed in a Jupyter notebook available at
\url{https://github.com/RuthAngus/stardate/blob/master/paper/code/Fitting_Praesepe.ipynb}.}.
This new Praesepe-calibrated gyrochronology relation is,
\begin{equation}
    \log_{10}(P_\mathrm{rot}) =
    c_A\log_{10}(t) +
    \sum_{n=0}^4 c_n[\log_{10}(G_{BP}-G_{RP})]^n
\label{eqn:fgk_gyro}
\end{equation}
for stars with \gcolor\ $<$ 2.7 and
\begin{equation}
    \log_{10}(P_\mathrm{rot}) =
    c_A\log_{10}(t) +
    \sum_{m=0}^1 b_m[\log_{10}(G_{BP}-G_{RP})]^m
\label{eqn:m_gyro}
\end{equation}
for stars with \gcolor\ $>$ 2.7, where $P_{\mathrm{rot}}$ is rotation period
in days and $t$ is age in years.
Best-fit coefficient values are shown in table \ref{tab:coefficients}.
\begin{table}[h!]
  \begin{center}
      \caption{Coefficient values for equations \ref{eqn:fgk_gyro} and
      \ref{eqn:m_gyro}.}
    \label{tab:coefficients}
    \begin{tabular}{l|c} 
      Coefficient & Value  \\
      \hline
      $c_A$ & 0.65 $\pm$ 0.05 \\
      $c_0$ & -4.7 $\pm$ 0.5 \\
      $c_1$ & 0.72 $\pm$ 0.05 \\
      $c_2$ & -4.9 $\pm$ 0.2 \\
      $c_3$ & 29 $\pm$ 2 \\
      $c_4$ & -38 $\pm$ 4 \\
      $b_0$ & 0.9 $\pm$ 0.5 \\
      $b_1$ & -13.6 $\pm$ 0.1 \\
    \end{tabular}
  \end{center}
\end{table}

We calibrated this new relation using Praesepe and the Sun alone, without
including other clusters, because different open clusters have slightly
different period-color relationships \citep{agueros2018, agueros2018b,
curtis2018, curtis2019} and, given that we used an age-color separable
relation, adding in extra clusters was unlikely to improve the calibration.
We did not include asteroseismic stars because most are slightly evolved and
would require a gyrochronology relation that depends on $\log g$.
This new relation, fit to Praesepe and the Sun, does not perfectly predict the
rotation periods of stars at all colors and ages, but it provides several
improvements over previous empirical gyrochronology relations.
Firstly, it uses new \ktwo\ rotation period measurements to model the
period-color relation of Praesepe in detail, secondly, it includes a model for
the rotational behavior of M dwarfs, and thirdly, it is calibrated to \Gaia\
\gcolor\ color: a directly observable quantity and the most widely available
photometric color index.

Equation \ref{eqn:fgk_gyro}, describing the rotational evolution of FGK and
early M stars, is most closely analogous to previously calibrated empirical
gyrochronology relations \citep[\eg][]{barnes2003, barnes2007, mamajek2008,
barnes2010, angus2015}.
It describes stars with Sun-like magnetic dynamos that follow a
`Skumanich-like' magnetic braking law, \ie\ their rotation period increases
with the square root of their age.
It does not describe stars hotter than around 6250 K which have a thin
convective layers and a weak magnetic dynamo, nor does it describe fully
convective stars which take a long time to converge onto the
\citet{skumanich1972} braking law \citep{krishnamurthi1997}.
In addition, stars with Rossby numbers larger than around 2 do not show
Skumanich-like magnetic braking \citep{vansaders2016, vansaders2018}, so
equation \ref{eqn:fgk_gyro} does not describe these stars.
It also does not describe the rotation periods of subgiants or giants, whose
rotation periods are influenced by their expanding radii, changing winds, and
core-to-surface differential rotation \citep[\eg][]{vansaders2013, tayar2018}.
Finally, this relation does not describe the rotation periods of dynamically
or magnetically interacting binaries which often rotate more rapidly than
isolated stars at the same age and color \citep{douglas2016}.
In order to include non-Skumanich type stars in our combined gyrochronal and
isochronal model, we designed a composite gyrochronology relation which
describes a mean and variance model for the rotation period distributions
across the HRD.
Rotation periods are modeled differently for stars of different photometric
color, EEP, Rossby number, age and metallicity.
The rotation period model for each of these groups is described below.
\begin{itemize}
    \item{The rotation periods of late F, GK, and early M dwarfs
        (0.56 $<$ \gcolor\ $<$ 2.7), with Rossby numbers less than 2, are
        modeled as a log-normal distribution, with mean given by equation
        \ref{eqn:fgk_gyro}, and variance given by the squared inverse of their
        observational uncertainties.}
    \item{The rotation periods of fully convective stars (\gcolor\ $>$ 2.7)
        are modeled with a log-normal distribution with mean given by equation
        \ref{eqn:m_gyro}, and an extra standard deviation of
        0.5 dex, added to observational uncertainties.
        This distribution reflects the observed rotation periods of late M
        dwarfs in the Praesepe cluster which span a broad range at every
        color.
        Late M dwarfs with masses $\lesssim$ 0.3 M$_\odot$, temperatures
        $\lesssim$ 3500 and \gcolor $\gtrsim$ 2.7 exhibit weak magnetic
        braking until at least after the age of Praesepe ($\sim$650 million
        years).
    \item{The rotation periods of F-type and hotter stars, with \gcolor\ $<$
        0.56 dex, are modeled as a log-normal distribution with mean,
        $\log_{10}(P_\mathrm{rot})=0.56$,
        and an additional standard deviation, added to observational
        uncertainties, of 0.5 dex.
        Stars more massive than around 1.25 M$_\odot$, with a temperature
        $\gtrsim$ 6250 K and a \gcolor\ $\lesssim$ 0.56 do not spin down
        appreciably over their main-sequence lifetimes because they do not
        have the deep convective envelope needed to generate strong magnetic
        fields.
        This model for the mean and variance of hot star rotation periods is
        based on stars hotter than 6250 K in the \citet{mcquillan2014} sample,
        which have a mean $\log_{10}(P_\mathrm{rot})$ of 0.56 dex and a
        standard deviation of around 0.5 dex.}
        Both hot and cool stars retain rotation periods that are similar to
        their primordial distribution \citep[see \eg][]{matt2012,
        somers2017}.}
    \item{The rotation periods of stars with large Rossby numbers are
        modeled as follows.
        The age at which a star's Rossby number would exceed 2 is calculated
        by inverting equation \ref{eqn:fgk_gyro}.
        If a star's age is greater than this, its mean rotation period is
        given by $P_\mathrm{max} = 2/\tau$, where $\tau$ is the convective
        turnover timescale, calculated via stellar mass using equation 11 from
        \citet{wright2011}.}
    \item{The rotation periods of subgiants are described with a log-normal
        distribution with mean given by equation \ref{eqn:fgk_gyro},
        \ref{eqn:m_gyro} or 0.56, depending on its color, and an additional
        standard deviation of 5 dex.
        This is not an accurate model of subgiant rotation
        periods \citep[see, \eg][]{vansaders2013} but the highly
        inflated variance makes it a weakly constraining one.
        Isochrone fitting provides precise ages for subgiants, so by inflating
        the variance of the gyrochronology relation at large EEP, we allow a
        star's position on the HRD/CMD to dominate the age information over
        its rotation period.
        This is useful because we have not yet built an accurate rotation-age
        relation for subgiants into our model.}
    \item{We model stars younger than around 250 Myrs with a log-normal
        distribution with mean function given by equation \ref{eqn:fgk_gyro},
        \ref{eqn:m_gyro} or 0.56, depending on it whether it has
        an intermediate, red, or blue color respectively and a inflated
        standard deviation of 0.5 dex.
        Rotation periods in young open clusters show a large amount of scatter
        because they have not yet converged onto the \citet{skumanich1972}
        spin-down sequence.}
    \item{Finally, we model stars with very low and high metallicites (-0.2
        $>$ [Fe/H] $>$ 0.2) with a log-normal distribution with mean given by
        equation \ref{eqn:fgk_gyro}, \ref{eqn:m_gyro} or 0.56, depending on
        its color, and a standard deviation of 0.5 dex.
        Gyrochronology is not calibrated at these extreme metallicities due to
        a lack of suitable metal poor and rich calibration stars.
        Rather than assume the same gyrochronology model can be na\"ively
        applied to these stars, we take a more conservative approach and model
        them with a broad Gaussian distribution.}
\end{itemize}
Inflating the variance of the rotation period distribtion for stars with
non-Skumanich magnetic braking behavior has two purposes: 1) in the cases of
hot stars and fully convective stars, it allows the broad distributions of
rotation periods observed in clusters and the field to be matched, and 2) it
down-weights the age-information provided by rotation periods in regions of
the HRD/CMD where rotation periods are not information-rich or the
gyrochronology model is inaccurate or uncalibrated.
If the observational uncertainties on rotation periods are 5\% on average,
which corresponds to 0.05/$\ln(10)$ $\sim$ 0.02 dex, adding 0.5 amounts to
a 25$\times$ increase in standard deviation, or a $\sim$ 600$\times$ increase
in variance.
So, practically speaking, when the standard deviation is inflated to 0.5 dex
or more, ages are almost entirely inferred via isochrone fitting.

\begin{figure}
  \caption{
    The additional rotation period scatter, $\sigma_P$, added to the
    observational period uncertainties in the model (see equation
    \ref{eqn:gyro}).
    The standard deviation was increased for early F and hotter
    stars (\gcolor $<$ 0.56), late M dwarfs (\gcolor\ $>$ 2.7) and evolved
    stars (EEP $\gtrsim$ 420) in order to down-weight the age-information
    supplied
    by rotation periods and reproduce observed rotation period distributions.
    We also increased the variance for stars younger than around 250 Myrs,
    because the rotation periods of these stars typically have not yet
    converged onto a tight gyrochronology sequence, and for very high and low
    metallicity stars (-0.2 $\gtrsim$ [Fe/H] $\gtrsim$ 0.2) because the
    gyrochronology relations have not been calibrated at these extreme values.
    Down-weighting the gyrochronal likelihood by the inverse variance
    ($1/\sigma^2$) allowed the ages of these stars to be mostly inferred
    via isochrone fitting.
    Sigmoid functions were used to provide smooth transitions between regions
    of low and high variance.
}
  \centering
    \includegraphics[width=1.\textwidth]{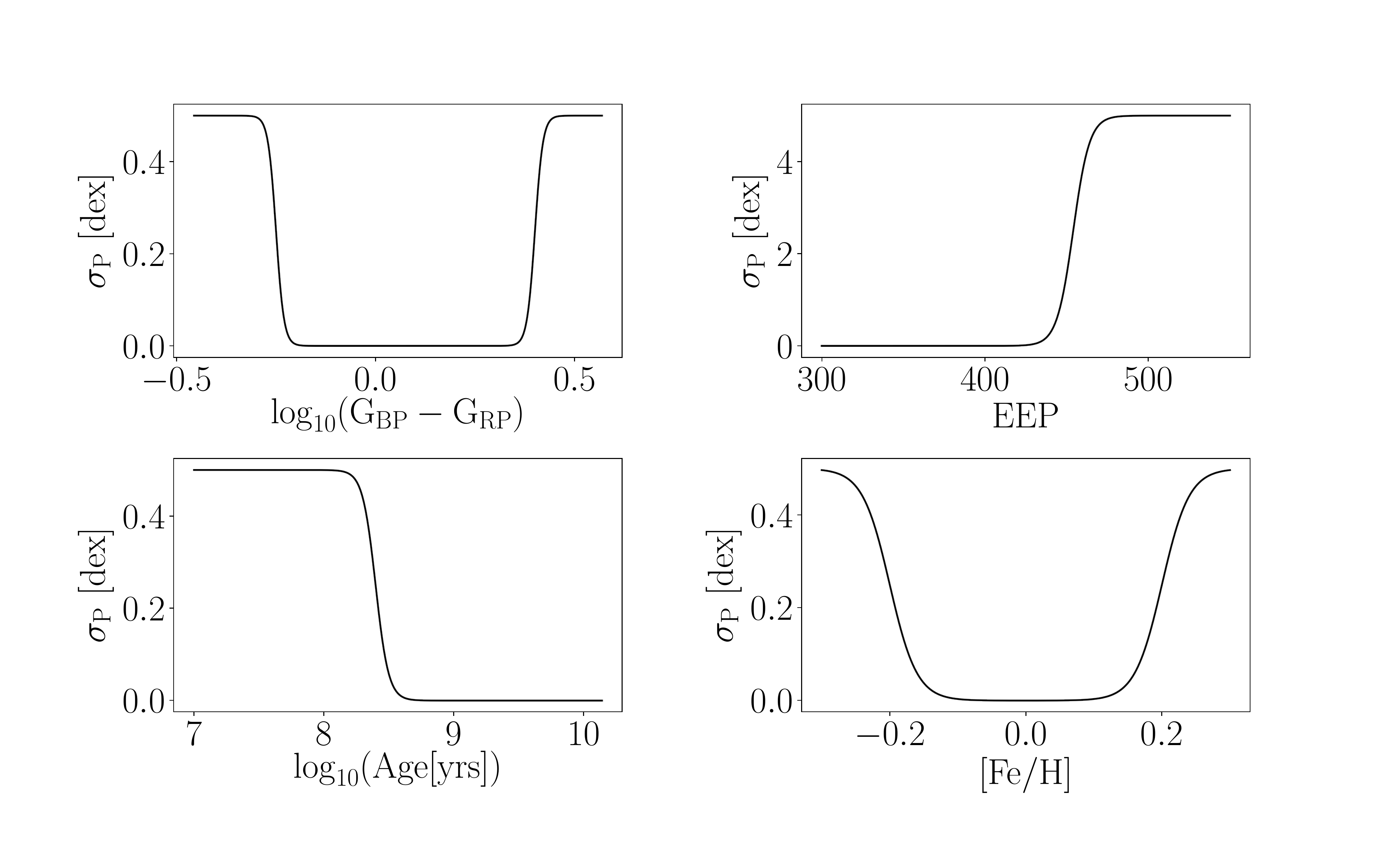}
\label{fig:variance}
\end{figure}

We used the following composite gyrochronology model to infer ages from
rotation periods,
\begin{equation}
    \log_{10}(P_\mathrm{rot}) \sim \begin{cases}
        \mathcal{N}\left[(\ref{eqn:fgk_gyro}), (\sigma+\sigma_P)^2\right], & Ro < 2,
        0.56 < G_{BP} - G_{RP} < 2.7 \\
        \mathcal{N}[(\ref{eqn:m_gyro}), (\sigma+\sigma_P)^2], & Ro < 2,
        G_{BP} - G_{RP} > 2.7 \\
        \mathcal{N}[\log_{10}(P_\mathrm{max}), (\sigma+\sigma_P)^2],& Ro \geq 2 \\
        \mathcal{N}[0.56, (\sigma+\sigma_P)^2],& G_{BP} - G_{RP} < 0.56,
    \end{cases}
\label{eqn:gyro}
\end{equation}
where $\sigma$ is the relative period uncertainty, divided by $\ln(10)$, on
individual rotation period measurements and $\sigma_P$ is an additional
scatter that is a function of EEP, age, metallicity and color.
It takes a maximal value of 0.5 for hot stars and fully convective stars, 5
for subgiants and giants, and a minimal value of zero for late F, GK and early
M dwarfs.
The variance model is shown in figure \ref{fig:variance}.
Sigmoid functions were used to provide smooth transitions between regions of
low and high variance.
Sharp changes in variance would produce sharp changes in likelihood, which
would cause the posterior distributions over stellar parameters to be more
difficult to sample.
The sigmoid functions shown in figure \ref{fig:variance} reach half their
maximum values at \gcolor\ = 0.56 and 2.5 for hot and cool stars respectively,
EEP = 454 for subgiants, age = 250 Myrs for young stars, $[$M/H$]$ = -0.2 for
metal poor stars and 0.2 for metal rich stars.
The logistic growth rate, or steepness, of the sigmoid functions is 100
dex$^{-1}$ for both color transitions, .2 EEP$^{-1}$ for the EEP transition,
20 dex$^{-1}$ for the age transition and 5 dex$^{-1}$ for the both metallicity
transitions.
The additional standard deviation is additive, so if a star is \eg\ hot,
evolved and metal poor, the additional standard deviation of its rotation
period rises to 6.

\subsection{Simultaneously fitting gyrochronology and isochrones}
The previous part of this section describes the model for the mean and
variance of rotation periods as a function of their ages (and colors, EEPs,
masses and metallicities).
In what follows, we describe how this model was combined with a stellar
evolution model to infer stellar ages (and other parameters) via
gyrochronology and isochrone fitting simultaneously.
Our goal was to infer the age of a star from its observable properties by
estimating the posterior probability density function (PDF) over age,
\begin{equation} \label{eqn:eqn1}
    p(t|{\bf m_x}, P_{\mathrm{rot}}, \bar{\pi}),
\end{equation}
where $t$ is age, ${\bf m_x}$ is a vector of
apparent magnitudes in various bandpasses,
\prot\ is the rotation period and \pmega\ is parallax.
Spectroscopic properties (\teff, \logg\ and $\mathrm{[Fe/H]}$) and/or
asteroseismic parameters (\dnu\ and \numax) may also be available for a star,
in which case they would appear to the right of the `$|$' in the above
equation since they are observables.
In order to calculate a posterior PDF over age, other stellar parameters must
be marginalized over.
These parameters are distance ($D$), V-band extinction ($A_V$), the
inferred metallicity, $[M/H]$\footnote{The inferred metallicity, [M/H] is a
model parameter which is different to the {\it observed} metallicity, [Fe/H]
which would appear on the right side of the $`|'$ in equation \ref{eqn:eqn1}.},
and equivalent evolutionary phase (abbreviated to EEP or
$E$).
EEP is a dimensionless number ranging from around 200 for M dwarfs up
to around 1600 for giants and is 355 for the Sun \citep[see][]{dotter2016,
choi2016}.
Stars are defined as subgiants when their EEP exceeds 454.
Mass is uniquely defined by EEP, age and metallicity.
The marginalization involves integrating over these extra parameters,
\begin{eqnarray} \label{eqn:bayes}
    & p(t|{\bf m_x},
    P_{\mathrm{rot}}, \bar{\pi})
\\ \nonumber
    & \propto \int p({\bf m_x},
    P_\mathrm{rot}, \bar{\pi}|
    t, E, [M/H], D, A_V)~p(t)\, p(E)\, p([M/H])\, p(D)\, p(A_V)\, \mathrm{d}E\,
    \mathrm{d}[M/H]\, \mathrm{d}D\, \mathrm{d}A_V.
\end{eqnarray}
This equation is a form of Bayes' rule,
\begin{equation} \label{eqn:eqn2}
\mathrm{Posterior} \propto \mathrm{Likelihood} \times \mathrm{Prior},
\end{equation}
where the likelihood of the data given the model is,
\begin{equation} \label{eqn:full_likelihood}
    p({\bf m_x}, \bar{\pi}, P_{\mathrm{rot}}|t, E, D, A_V, [M/H]),
\end{equation}
and the prior PDF over parameters is,
\begin{equation} \label{eqn:prior}
    p(t)\, p(E)\, p(D)\, p(A_V)\, p([M/H]).
\end{equation}
The priors we used are described in the appendix.

We assumed that the process of magnetic braking is independent of hydrogen
burning in the core, outside of the dependencies that are captured in the
model.
This assumption allowed us to multiply two separate likelihood
functions together: one computed using an isochronal model and one computed
using a gyrochronal model.
We assumed that the probability of observing the measured observables given
the model parameters was a Gaussian and that the observables were identically
and independently distributed.
The isochronal likelihood function was,
\begin{eqnarray} \label{eqn:isochrones_only_likelihood}
    & \mathcal{L_{\mathrm{iso}}} = p({\bf m_x},
    \bar{\pi}|t, E, [M/H], D,
    A_V) \\ \nonumber
    & = \frac{1}{\sqrt{(2\pi)^n \det(\Sigma)}}
    \exp\left( -\frac{1}{2} ({\bf O_I} - {\bf I})^T \Sigma ^{-1}
    ({\bf O_I} - {\bf I})\right),
\end{eqnarray}
where ${\bf O_I}$ is the vector of n observables: \pmega, ${\bf m_x}$ plus
spectroscopic and/or asteroseismic observables if available, and $\Sigma$ is
the covariance matrix of that set of observables.
${\bf I}$ is the vector of {\it model} observables that correspond to a set of
parameters: $t$, $E$, $[M/H]$, $D$ and $A_V$, calculated using an isochrone model.
We assumed there is no covariance between these observables and so this
covariance matrix consists of individual parameter variances along the
diagonal with zeros everywhere else.
The gyrochronal likelihood function was,
\begin{eqnarray} \label{eqn:gyro_likelihood}
    & \mathcal{L_{\mathrm{gyro}}} = p(P_\mathrm{rot} |t, E, [M/H], D, A_V) \\ \nonumber
    & = \frac{1}{\sqrt{(2\pi) \det(\Sigma_P)}}
    \exp\left( -\frac{1}{2} ({\bf P_O} - {\bf P_P})^T \Sigma_P ^{-1}
    ({\bf P_O} - {\bf P_P})\right),
\end{eqnarray}
where ${\bf P_O}$ is a 1-D vector of observed logarithmic rotation periods,
and ${\bf P_P}$ is the vector of corresponding logarithmic rotation periods,
predicted by the model.
$\Sigma_P$ was comprised of individual rotation period measurement
uncertainties, plus an additional variance that is a function of EEP and
\gcolor\
color, added in quadrature.
This variance accounts for the stochastic nature of the rotation periods of
very hot and very cool stars and allowed us to predominantly use isochrone
fitting to measure the ages of subgiants.
The full likelihood used in our model was the product of these two likelihood
functions,
\begin{equation}
    \mathcal{L}_{\mathrm{full}} = \mathcal{L}_{\mathrm{iso}} \times
    \mathcal{L}_{\mathrm{gyro}}.
\end{equation}

The inference processes proceded as follows.
First, a set of parameters: age, EEP, metallicity, distance and extinction,
and observables for a single star were passed to the isochronal likelihood
function in equation \eqref{eqn:isochrones_only_likelihood}.
Then, a set of observables corresponding to those parameters were generated
from the MIST model grid using {\tt isochrones.py} \citep{isochrones} and
compared to the measured observables via the isochronal likelihood,
$\mathcal{L}_{\mathrm{iso}}$ (also computed using {\tt isochrones.py}).
The parameters were also passed to the gyrochronology model (equation
\ref{eqn:gyro}) where $t$, $E$,
$[M/H]$, $D$ and $A_V$ were used to calculate \gcolor\ color and mass from the
MIST model grid and, in turn, rotation period via the gyrochronology model.
EEP and \gcolor\ were also used to calculate the additional rotation period
variance, added to the individual period uncertainties.
This model rotation period was compared to the measured rotation period using
gyrochronal likelhood function of equation \ref{eqn:gyro_likelihood}.
The gyrochronal log-likelihood was added to the isochronal log-likelihood to
give the full likelihood, which was then added to the log-prior to produce a
single sample from the posterior PDF.

Ages were inferred with \sd\ using Markov Chain Monte Carlo.
The joint posterior PDF over age, mass, metallicity, distance and extinction
was sampled using the affine invariant ensemble sampler, {\tt emcee}
\citep{foreman-mackey2013} with 50 walkers.
Samples were drawn from the posterior PDF until 100 {\it independent} samples
were obtained.
We actively estimated the autocorrelation length, which indicates how many
steps were taken per independent sample, after every 100 steps using the
autocorrelation tool built into {\tt emcee}.
The MCMC concluded when {\it either} 100 times the autocorrelation length was
reached and the change in autocorrelation length over 100 samples was less
than 0.01, {\it or} the maximum of 500,000 samples was obtained.
This method is trivially parallelizable, since the inference process for each
star can be performed on a separate core.
The age of a single star can be inferred in around 1 hour on a laptop
computer.

\section{Results}
\label{section:results}

In order to demonstrate the performance of our method, we conducted three sets
of tests.
In the first we simulated observables from a set of stellar parameters for a
few hundred stars using the MIST stellar evolution models and the
gyrochronology model of equation \ref{eqn:gyro}.
The ages predicted with our model were compared to the true parameters used to
generate the data.
In the second we tested our model by measuring the ages of individual stars in
the NGC 6819 open cluster, and in the third we tested our model on \kepler\
asteroseismic stars.

\subsection{Test 1: simulated stars}
\begin{figure}
  \caption{
      The simulated star sample plotted on an HRD, colored by age
    (top panel) and rotation period (bottom panel).
    HRD positions were calculated using MIST isochrones via the {\tt
    isochrones.py} {\it Python} package and rotation periods were generated
    using equation \ref{eqn:gyro}.
    This figure was generated in a Jupyter notebook available at
    \url{https://github.com/RuthAngus/stardate/blob/master/paper/code/Simulate_data.ipynb}
}
  \centering
    \includegraphics[width=1\textwidth]{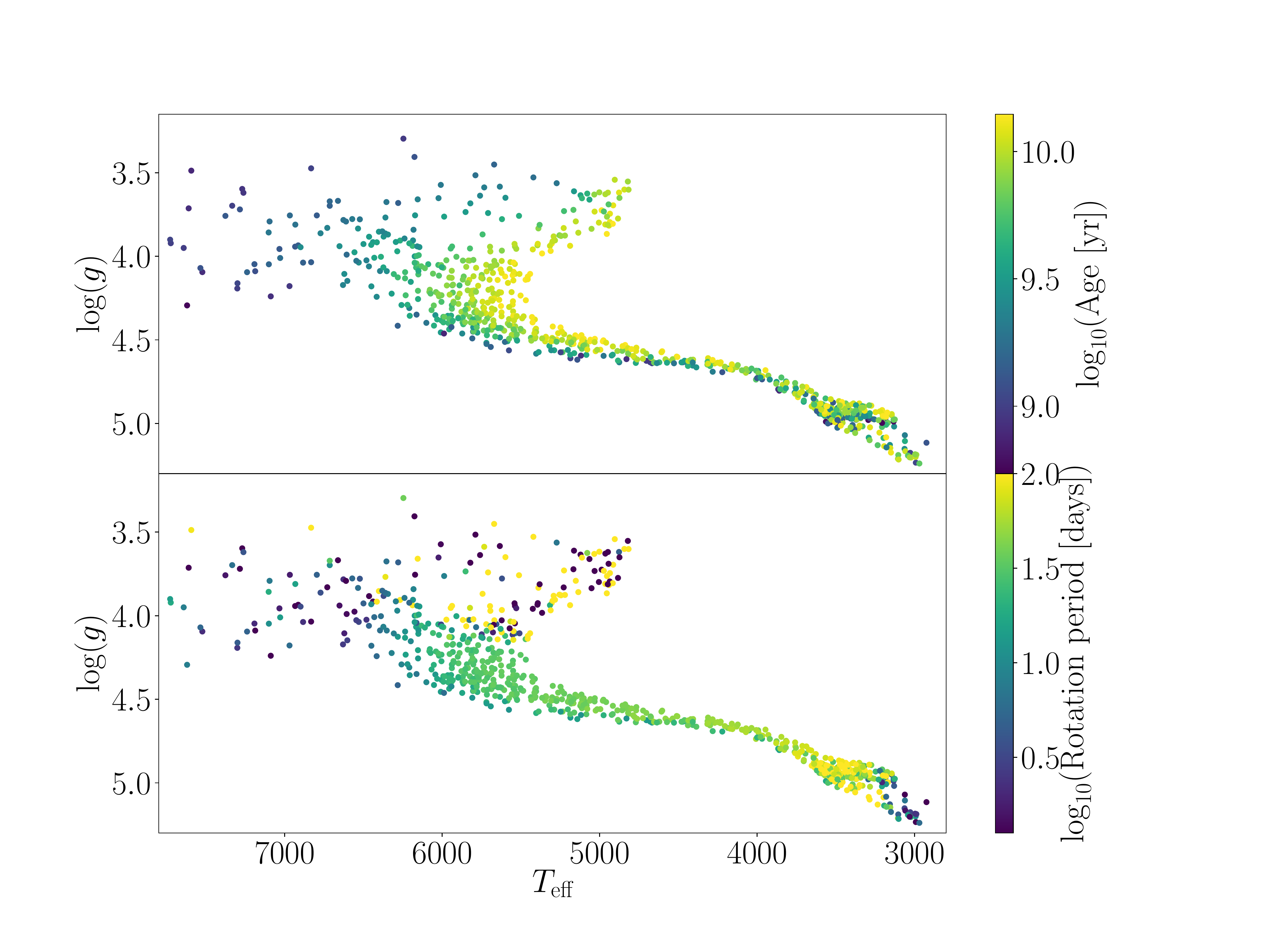}
\label{fig:CMD_age}
\end{figure}

\begin{figure}
  \caption{
Data simulated from the rotation period model.
    Late F, GK and early M dwarfs (stars with 0.56 $<$ \gcolor\ $<$ 2.7 follow
    the Praesepe-calibrated gyrochronology relation (dashed gray lines), with
    the exception of old, slowly
    rotating stars with large Rossby numbers whose rotation periods are fixed
    at 2$\times$ their convective overturn time.
    The rotation periods of early F (\gcolor $<$ 0.56), late M dwarfs (\gcolor
    $<$ 2.7) and subgiants (EEP $\gtrsim$ 420) were generated
    from a log-normal distribution with standard deviation given by equation
    \ref{eqn:gyro}.
The top panel shows the rotation periods vs. \gcolor\ colors of simulated stars,
    colored by their age and the bottom panel shows the same stars colored
    by their equivalent evolutionary phase (EEP).
    The gray lines describe the mean gyrochronology model at ages 1,
    3, 5, 7, 9, 11, and 13 (rotation periods rise with age).
    This figure was generated in a Jupyter Notebook, available at
    \url{https://github.com/RuthAngus/stardate/blob/master/paper/code/Simulate_data.ipynb}
}
  \centering
    \includegraphics[width=1.\textwidth]{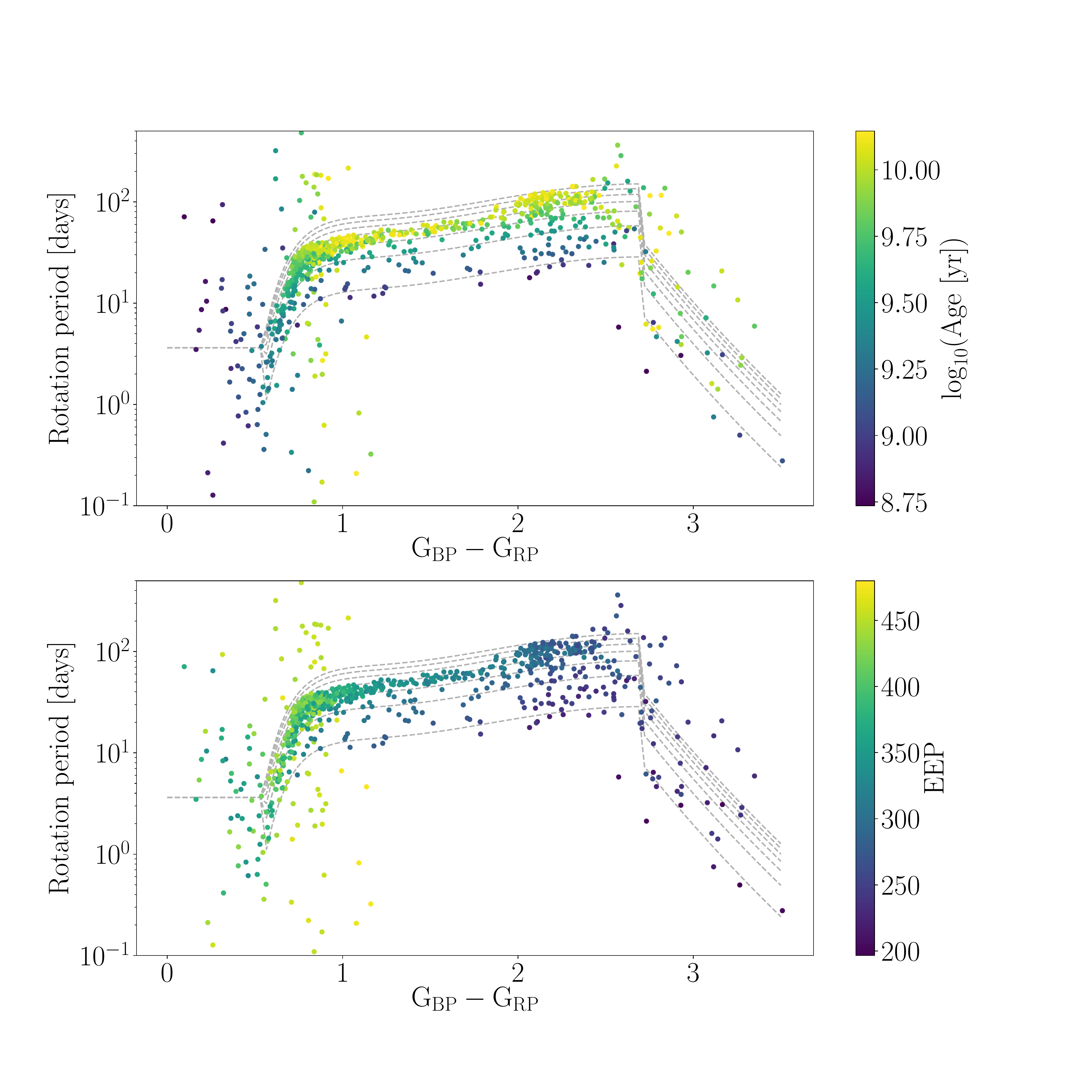}
\label{fig:rotation_model}
\end{figure}

For the first test we drew masses, ages, bulk metallicities, distances and
extinctions at random for 1000 stars from the following uniform distributions:
\begin{eqnarray}
& \mathrm{EEP} \sim U(198, 480) \\
& t \sim U(0.5, 14)\mathrm{~[Gyr]} \\
& [Fe/H] \sim U(-0.2, 0.2) \\
& D \sim U(10, 1000)~\mathrm{[pc]} \\
& A_V \sim U(0, 0.1).
\end{eqnarray}
\teff, \logg, [Fe/H], parallax, and apparent magnitudes $B$, $V$, $J$, $H$, $K$,
\gaia\ $G$, $G_{BP}$ and $G_{RP}$ were generated from these
stellar parameters using the MIST stellar evolution models.
We added a small amount of noise to the `observed' stellar properties in order
to reflect optimistic observational uncertainties for isochrone-dating.
We added Gaussian noise with a standard deviation of 25 K to \teff, 0.01 dex
to \feh\ and \logg, and 10 mmags to $B$, $V$, $J$, $H$, and $K$ magnitudes.
These are just one choice of uncertainties that we could have adopted and are
extremely optimistic.
The uncertainties on predicted ages {\it will} depend strongly on all
observational uncertainties, however, since this analysis is designed to show
the relative improvement in stellar age precision when {\it rotation periods}
are included, we chose to use best-case spectroscopic parameters.
The noise added to Gaia $G$-band photometry ranged from
0.3 mmag for stars brighter than 13th magnitude, to 10 mmag for stars
around 20th magnitude \citep{evans2017, brown2018}.
Noise added to \gaia\ $G_{BP}$ and $G_{RP}$ bands ranged from 2 mmag for stars
brighter than 13th magnitude to 200 mmag for stars fainter than 17th.
Unphysical combinations of stellar parameters were discarded, resulting in a
final sample size of 841 simulated stars.
Figure \ref{fig:CMD_age} shows the position of these stars on an HRD
(with \logg\ on the y-axis instead of luminosity to improve the visibility of
the MS), colored by their age.
Rotation periods for these stars were generated using the gyrochronology
relation described in equation \ref{eqn:gyro}.
We added 5\% Gaussian noise to all stellar rotation periods to represent
realistic measurement uncertainties of 5\%.
The median uncertainty on rotation periods calculated from \kepler\ light
curves, provided in the \citet{mcquillan2014} catalog is 1\%.
However, the \citet{aigrain2015} injection and recovery study showed that true
rotation period uncertainties are often slightly larger than this, and noise
distribution of rotation periods can be highly non-Gaussian
\citep[\eg][]{aigrain2015, angus2018}.
Figure \ref{fig:rotation_model} shows the rotation periods of 841
stars generated from the gyrochronology model.

\begin{figure}
  \caption{
The true vs. predicted ages of simulated stars.
    Ages calculated by combining gyrochronology
    and isochrone fitting with \sd\ are shown in color and ages calculated with
    isochrone fitting only are shown in gray.
The different panels show the results for stars
    with \gcolor\ $<$ 2.2 (FGK dwarfs) that are still braking magnetically
    ($Ro < 2$), stars with \gcolor\ $<$ 2.2 that have stopped braking
    magnetically ($Ro \geq 2$),
    stars with 2.2 $<$ \gcolor\ (M dwarfs), and evolved stars (EEP $>$ 420).
Gyrochronology is highly effective for FGK stars and ages inferred with both
    gyrochronology and isochrone fitting are more accurate and precise than
    ages inferred via isochrone fitting only for this group.
Neither gyrochronology nor isochrone fitting can provide precise ages for
    M dwarfs, so the ages of these stars are imprecise regardless of
    age-dating method.
    This figure was generated in a Jupyter Notebook available at
    \url{https://github.com/RuthAngus/stardate/blob/master/paper/code/Results_plots.ipynb}.
}
  \centering
    \includegraphics[width=1\textwidth]{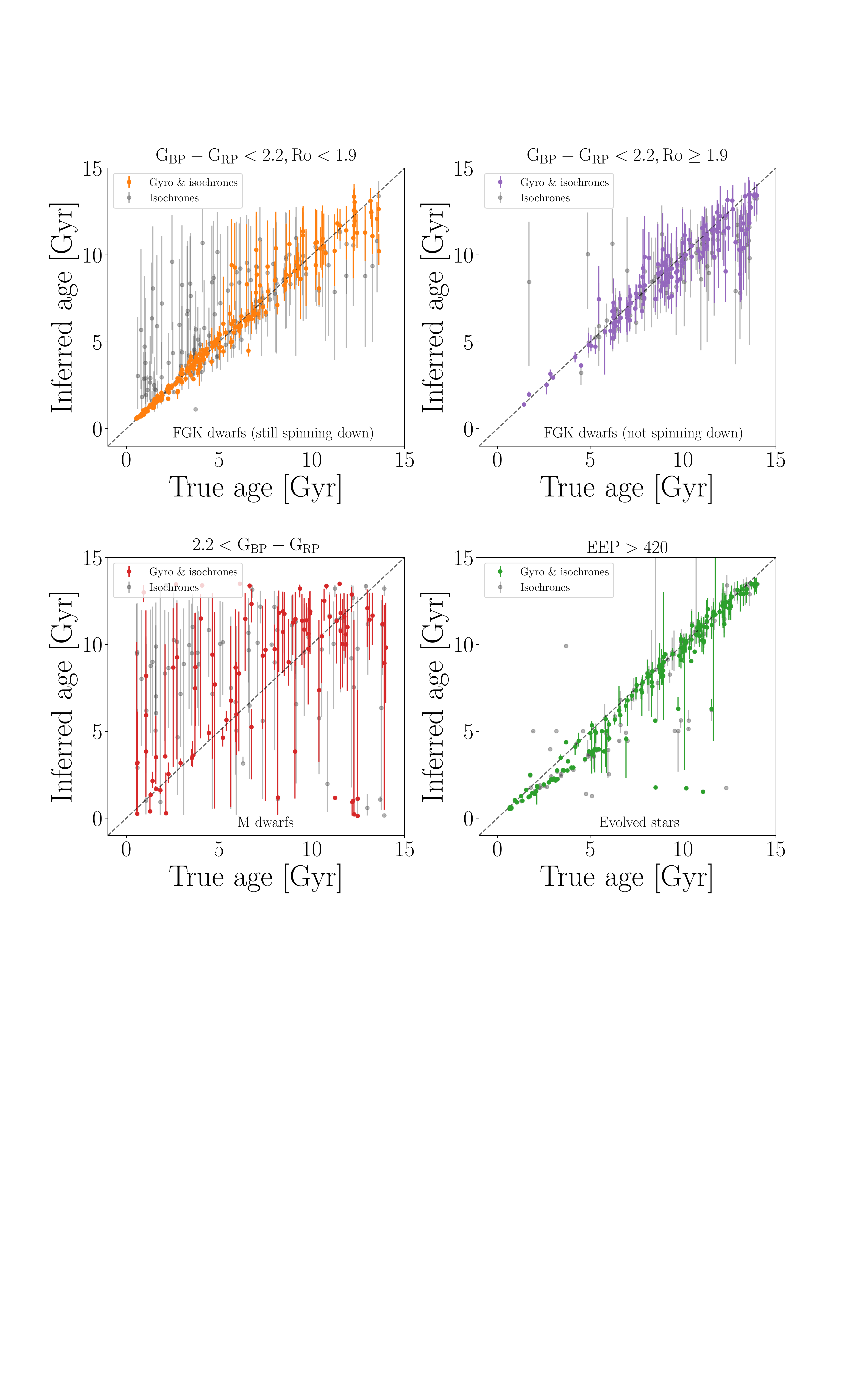}
\label{fig:simulation_results}
\end{figure}

We took two approaches to inferring the ages of these simulated stars:
firstly using isochrone fitting {\it only}, and secondly using isochrone
fitting {\it combined with} a gyrochronology model (\sd).
Since the posterior PDFs of stars are often multimodal, we found that the
choice of initial positions of the {\tt emcee} walkers influenced the final
outcome because walkers occasionally got stuck in local minima.
We found that the following set of initial parameters worked well, though not
perfectly: EEP = 330, $t = 9.56$ Gyr, $[M/H] = -0.05$, $D = 269$ pc and $A_V =
0.0$\footnote{These are the default initial parameters provided in \sd.}.
Figure \ref{fig:simulation_results} shows the results of combining
gyrochronology with isochrone fitting for the simulated sample.
The stars' true ages are plotted against their predicted ages, with ages
inferred with gyrochronology and isochrone fitting
in color, and ages inferred using isochrone fitting only plotted in light
grey.
The different panels show the results for different types of stars: FGK
dwarfs that are still undergoing magnetic braking, FGK dwarfs that have ceased
magnetic braking (their Rossby number is around 2), M dwarfs, and evolved
stars.
The selection criteria for these groups are in the panel headings.
The FGK dwarfs with low Rossby numbers showed the largest improvement: the
median age precision for this group (defined as the standard deviation of the
posterior as a percentage of the median age) was 8\% when using a combination
of isochrones and gyrochronology, and 22\% using isochrone fitting alone.
This equates to an almost 3$\times$ improvement in age precision.
The age RMS of this group was 0.8 Gyr using isochrones and gyrochronology, and
2 Gyr using isochrones only.
Despite the fact that stars with Rossby numbers of 2 have stopped spinning
down and their rotation periods no longer evolve with age, the ages of these
stars can still be relatively precisely constrained with isochrone fitting.
The median age precision for stars with large Rossby numbers was 11\% with
gyrochronology and isochrone fitting, and 13\% with isochrone fitting only.
The precision of M dwarf ages did improve overall when their rotation periods
were included, but this improvement was entirely driven by the early M dwarfs.
The precision of this group improved overall from 33\% to 22\% and RMS from
5.4 Gyr to 4.6 Gyr.
The precision of ages inferred for evolved stars changed very little when
rotation periods were included in the inference process.
This is because the variance of the rotation-age relation was inflated by a
large amount in the gyrochronology model (equation \ref{eqn:gyro}), making
rotation periods almost entirely uninformative for this group.
The median age precision of subgiants from both gyrochronology and isochrone
fitting and isochrone fitting alone was 7\%.

An important caveat associated with these results is that they strongly depend
on the uncertainties adopted for all observables used in the analysis: \teff,
\feh, \logg, $G$, $G_{BP}$, $G_{RP}$, $J$, $H$, $K$, $B$, $V$, and rotation
period.
Changing the uncertainties on these observables will affect the uncertainties
on inferred ages in different ways.
This demonstration is not intended to reflect the typical age uncertainties
that will result for all stars in reality, it merely exemplifies the increase
in stellar age precision that results from one specific choice of
uncertainties.
Needless to say, estimating the uncertainties on observables accurately can be
as important as accurately measuring the observables themselves.

This simulation experiment was designed to show the theoretical improvement in
age precision when gyrochronology is incorporated into isochrone fitting.
However, it does not demonstrate the accuracy of this method because the test
data were simulated from the same model used to infer ages.
The results of this experiment are therefore extremely accurate by design.
When applying this method to real data, the results will only be accurate if
the model is accurate.
In other words, \sd, like any age-dating method, provides model-dependent
ages.
Stellar ages calculated with \sd\ depend on both the accuracy of the MIST
models {\it and} the accuracy of the gyrochronology model (equation
\ref{eqn:gyro}).
In order to test the accuracy of \sd, we applied it to real data, as described
in the following section.

\subsection{Test 2: Open clusters}
\begin{figure}
  \caption{
    The \kepler-based rotation periods of members of the 2.5 Gyr NGC 6819 open
    cluster.
    The raw \gcolor\ colors are shown in red and the dust-corrected colors are
    shown in black.
    The dashed line shows a gyrochronology model that was fit to the Praesepe
    cluster and the Sun in this work, interpolated to 2.5 Gyrs.
    The solid blue line shows a previously calibrated gyrochronology model
    \citep{angus2015}.
    This figure was generated in a Jupyter Notebook available at:
    \url{https://github.com/RuthAngus/stardate/blob/master/paper/code/NGC6819.ipynb}
  }
  \centering
    \includegraphics[width=1\textwidth]{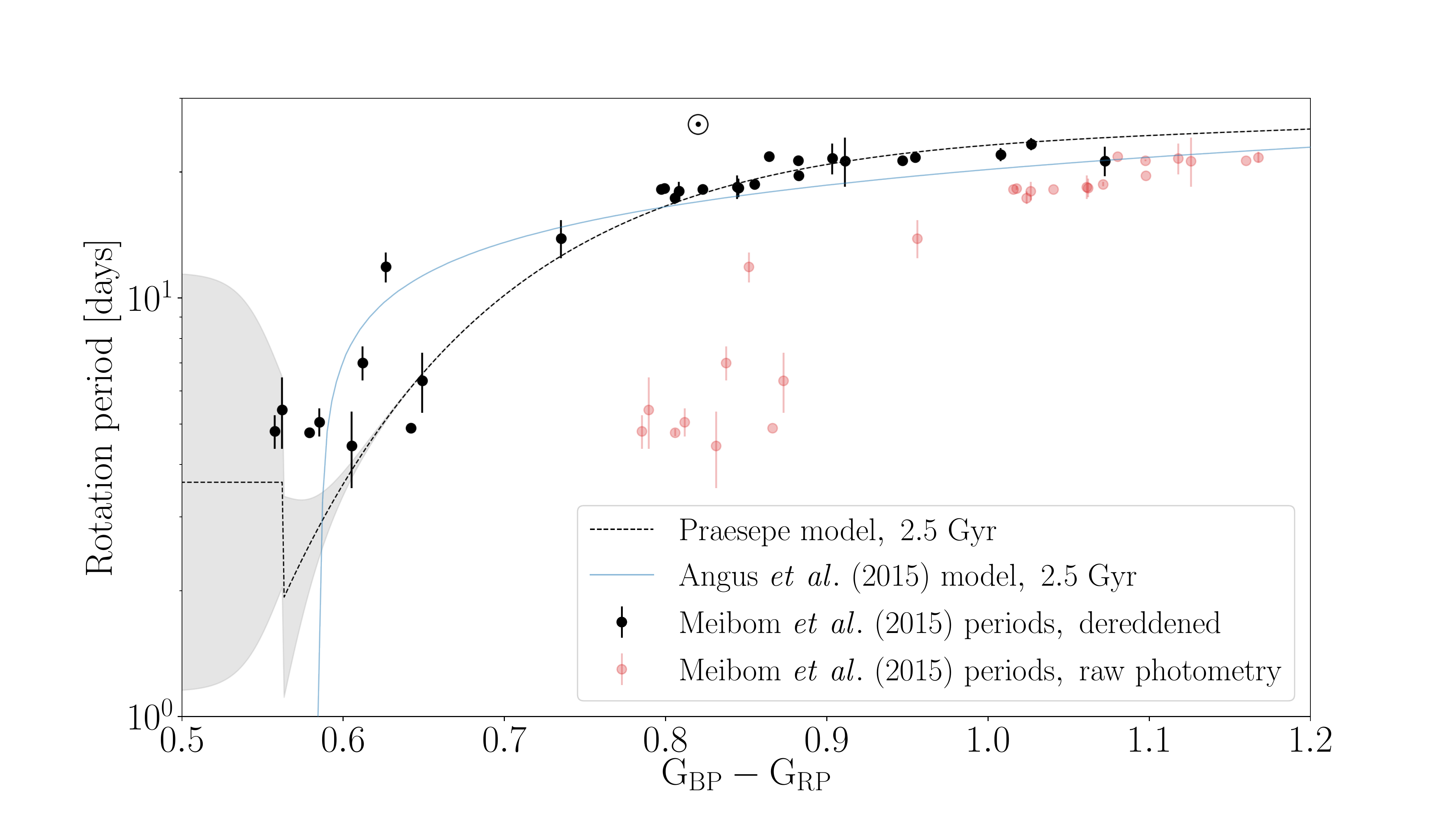}
\label{fig:NGC6819}
\end{figure}

\begin{figure}
  \caption{
    The inferred ages of members of the NGC 6819 open cluster as a function of
    their \gcolor\ color.
    Ages of stars inferred using a combination of isochrone fitting and
    gyrochronology (Praesepe and Sun calibration) with dereddened \gaia\ $G$,
    $G_{BP}$, and $G_{RP}$ photometry (black circles) and uncorrected, raw,
    photometry (red squares).
    Even though V-band extinction is marginalized over in the inference
    process, reddening can still bias ages.
    Blue triangles, pointing up, show ages inferred using isochrone fitting
    and gyrochronology, with the \citet{angus2015} gyrochronology model.
    Orange triangles, pointing down, show ages inferred using isochrone
    fitting only.
    The ages of F stars (stars bluer than 0.7) were precisely constrained by
    isochrones and including gyrochronology makes little difference to their
    inferred ages.
    The age precision of G and K dwarfs (stars redder than 0.7) was improved
    by including gyrochronology.
    The median age of stars inferred using the gyrochronology model
    calibrated to Praesepe and the Sun (black circles) was 2.65 $\pm$ 0.13
    which is consistent with the established cluster age (2.5 Gyr).
    This figure was generated in a Jupyter Notebook available at:
    \url{https://github.com/RuthAngus/stardate/blob/master/paper/code/NGC6819.ipynb}
}
  \centering
    \includegraphics[width=1\textwidth]{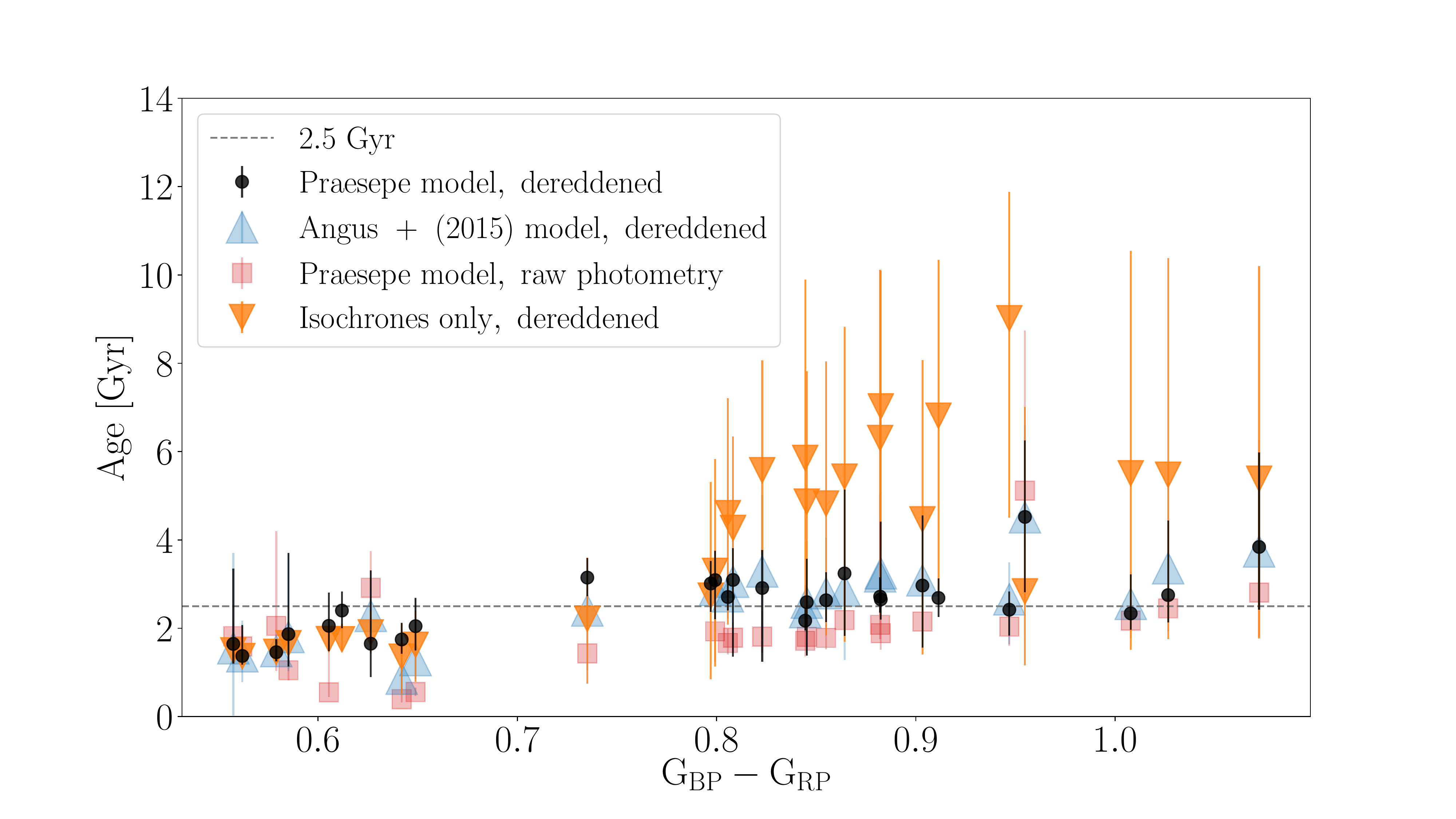}
\label{fig:NGC6819_results}
\end{figure}

In order to test our model on real stars with known ages, we selected a sample
of stars in the 2.5 Gyr NGC 6819 cluster.
We compiled \kepler-based rotation periods \citep{meibom2015}, \Gaia\
photometry and \gaia\ parallaxes for members of the NGC 6819 cluster.
Figure \ref{fig:NGC6819} shows the period-color relation of this
cluster and figure \ref{fig:NGC6819_results} shows the results of inferring
the ages of individual cluster members using a combination of gyrochronology
and isochrone fitting (via \sd) and isochrone fitting alone.
The ages of F stars in this cluster (\gcolor\ $\sim$ 0.5-0.75) were relatively
precisely constrained by isochrone fitting alone because, at 2.5 Gyr, they are
approaching the MS turnoff.
For these hot stars, ages inferred with gyrochronology and isochrones were
similar to ages inferred with isochrones and similarly precise, showing that
isochrones provide a lot of age information for these stars and rotation
periods do not add significantly more information.
The G and early K stars in this cluster (\gcolor\ $\lesssim$ .65) were not
precisely recovered from isochrone fitting alone -- the isochrone-only age
posteriors tend towards the prior which is a uniform distribution between 0
and 13.8 Gyrs.
The median age of stars in the cluster was 4.27 $\pm$ 0.48 Gyr when only
isochrone fitting was used.
In contrast, including gyrochronology when inferring the ages of G and K stars
in this cluster significantly improved age precision.
The median age of stars in this cluster was 2.63 $\pm$ 0.16 Gyr when ages were
inferred with a combination of isochrone fitting and gyrochronology, using the
newly calibrated Praesepe-based gyrochronology model.
The previously-calibrated \citet{angus2015} model resulted in a median stellar
age of 2.66 $\pm$ 0.21 Gyr, which is still consistent with the established
cluster age of 2.5 Gyr.
The median age of stars in the cluster using uncorrected photometry was
slightly underestimated at 1.86 $\pm$ 0.22 Gyr.
This suggests that, despite the fact that V-band extinction is marginalized
over during the inference process, correcting for extinction {\it before} ages
are estimated will reduce bias introduced by dust.

We found that 5\% rotation period uncertainties resulted in the most accurate
ages for NGC 6819.
The uncertainties on the measured rotation periods, provided in
\citet{meibom2015} and shown in figure \ref{fig:NGC6819}, were likely
underestimated for some stars.
Underestimated rotation period uncertainties can result in inaccurate age
estimates.
This raises the question: how should uncertainties on rotation periods be
estimated?
The likelihood is weighted by the inverse variance, so uncertainties on the
rotation period control the relative information provided by gyrochronology,
isochrones, and the prior.
If rotation period uncertainties are either too large or too small, the
resulting age estimate will be imprecise and/or inaccurate.
It is difficult to measure uncertainties on rotation periods directly:
standard techniques such as Lomb-Scargle periodograms and autocorrelation
functions do not provide them.
    Ideally, rotation period uncertainties should capture {\it both} the
measurement precision, {\it and} the physical uncertainty introduced by the
latitudinal movement of star spots on the surface of a differentially rotating
star.
For example, \citet{donahue1996} demonstrated that the seasonal variation in
measurements of G and K star rotation periods is a function of period, $\Delta
\mathrm{P_{rot}} \propto \mathrm{P_{rot}}^{1.3\pm0.1}$.
This variation is presumably caused by a latitudinal drift in the dominant
active regions, which traces the stellar cycle over several years, in
combination with latitudinal differential rotation.
This suggests that latitudinal spot drifting does not significantly
affect stellar rotation periods when stars are young, for example the scatter
of rotation periods about the mean gyrochronology model in Praesepe is only
around 5\%.
However it is likely that this effect will become more important at older
ages.
A thorough exploration of how rotation period uncertainties, from both
measurement uncertainty and physical variation, affect stellar ages via
gyrochronology is key to understanding the power of gyrochronology as an
age-dating method.
For now, we leave this exploration for a future study.

\subsection{Test 3: Kepler asteroseismic stars}
\begin{figure}
    \caption{ A comparison of stellar ages inferred using
    asteroseismic modeling with ages inferred using a combination of isochrone
    fitting and gyrochronology.
Colored circles show ages inferred using isochrone fitting and gyrochronology
combined via the \sd\ software package.
Black triangles show the ages of all stars inferred via gyrochronology only
    and white circles show ages inferred via isochrone-fitting only.
  }
  \centering
    \includegraphics[width=1.2\textwidth]{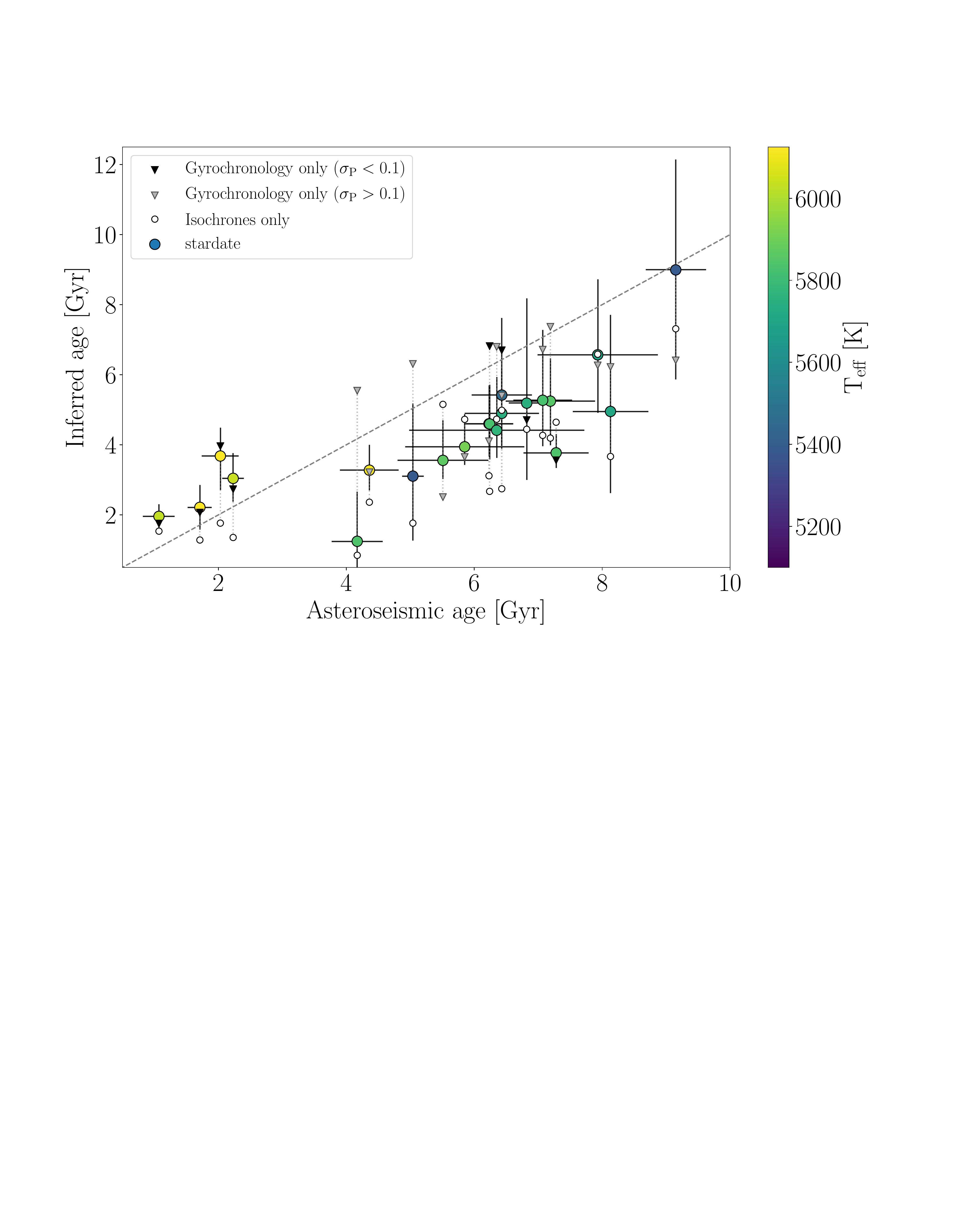}
\label{fig:astero}
\end{figure}

    In order to test our method in the regime where both isochrone
fitting and gyrochronology become important, we recovered the ages of the 21
asteroseismic stars analysed in \citet{vansaders2016}.
These 21 stars were observed in \kepler's short cadence mode and are a mixture
of dwarfs and subgiants.
Their asteroseismic ages were calculated from the analysis of the frequencies
of individual oscillation modes \citep{mathur2012, metcalfe2014,
silvaaguirre2015, ceillier2016}, and their rotation periods from their
\kepler\ light curves \citep{garcia2014}.
We crossmatched these stars with the \Gaia\ catalog to obtain parallaxes and
apparant magnitudes in the \Gaia\ G, $\mathrm{G_{BP}}$ and $\mathrm{G_{RP}}$
band passes.
We also added J, H and K 2MASS magnitudes from the Kepler input catalog
\citep{brown2011},
and used spectroscopic effective temperatures, spectroscopic metallicities and
rotation periods reported in table 1 of \citet{vansaders2016}.
The \gaia\ photometry is extremely precise, and we found that artificially
inflating the uncertainties on \gaia\ apparent magnitudes by a factor of 10
substantially improved the quality of fit, both in terms of MCMC convergence
and agreement with asteroseismic age measurements.
In figure \ref{fig:astero} we show the ages of these 21 stars inferred using
isochrone fitting and gyrochronology, against their asteroseismic ages,
calculated using the Asteroseismic Modeling Portal (AMP) \citep{metcalfe2009,
metcalfe2012, metcalfe2014}.
In this figure, the colored symbols show ages inferred using a combination of
gyrochronology and isochrone fitting, implemented with the \sd\ {\it Python}
package.
The black and grey triangles show ages inferred from gyrochronology only,
where the mass and color of the stars were not inferred, but fixed to be the
asteroseismic mass and the \gaia\ \gcolor\ color.
These gyrochronal ages were calculated with age as the only free parameter,
without marginalizing over stellar mass, \gcolor\ color or extinction.
As a result, these gyrochronal ages are {\it not the same} as the ages given
by the maximum of the gyrochronal likelihood function used in the combined
age model, they simply represent an approximation to the gyrochronal age.
Grey triangles are shown for stars where gyrochronology is not applicable
because the stars are either too metal poor, too metal rich, or too evolved.
Black triangles are shown for stars where gyrochronology {\it is} applicable.
The white circles show ages inferred from isochrones only.
Dashed lines connect the three different age measurements for the same
stars.

Although the ages of all 21 stars shown in figure \ref{fig:astero} were
inferred with a joint isochronal and gyrochronal model, most (all but 8) were
either too evolved, too metal poor, or too metal rich for gyrochronology to
contribute any information to the ages.
These metal poor/rich or evolved stars lie in a regime where the variance on
their rotation period was artificially inflated because the gyrochronology
relations are not thoroughly understood or well calibrated.
The rotation periods of the remaining eight stars {\it did} contribute to
their inferred ages to some degree, however isochrones still dominated the age
information for some of them because these stars are relatively old and/or
relatively hot.

In general, there is relatively poor agreement between the asteroseismic ages
and the ages inferred using \sd.
Much of this discrepancy is driven by differences in the isochronal ages,
which is likely attributable to differences between the MIST stellar evolution
models and those used in the AMP analysis: a combination of the Aarhus stellar
evolution code \citep[ASTEC][]{christensen-dalsgaard2008a} and the adiabatic
pulsation code \citep[ADIPLS][]{christensen-dalsgaard2008b}.
We compared non-rotating, Solar-metallicity MIST isochrones for middle-aged
stars with Solar-metallicity BaSTI isochrones \citep{pietrinferni2004,
hidalgo2018} and found that, for stars between 4 and 8 Gyrs, in the same
effective temperature range as the asteroseismic stars, the age discrepancy
between the two sets of models can be as large as 1-2 billion years.
The MIST isochrones lie above the BaSTI isochrones on the HR diagram, leading
to a systematic underprediction of ages.

The gyrochronal ages, where gyrochronology is applicable, do not show
excellent agreement with the asteroseismic ages either.
The four hot stars to the left in figure \ref{fig:astero} are rotating more
slowly than predicted by the Praesepe-based gyrochronology models, and as a
result their gyro-ages are older than their asteroseismic ones.
For these hot stars, two out of four have ages that are still consistent, or
close to consistent, with their asteroseismic ages.
The third star from the left is an anomalously slow rotator for its age and
mass and \citet{vansaders2016} also found this star to be surprisingly slowly
rotating.
In contrast, the star with a black triangle symbol (indicating that
gyrochronology is applicable) furthest to the right is {\it rapidly} rotating
for its mass and age, even when weakened braking is taken into account.
This star is around Solar mass (1.0 $\pm$ 0.03 M$_\odot$) and has an
asteroseismic age older than the Sun (7.28 $\pm$ 0.51 Gyr), yet rotates with a
period of only 19.8 $\pm$ 1.3 days.
This is KIC 9098294, a single-lined spectroscopic binary with an orbital
period of around 20 days (Latham, private communication).
It is the only clear SB1 in the \citet{vansaders2016} sample, although some
others do have binary companions with long orbital periods, for which tidal
interactions are not expected to be strong.

\section{Conclusions}
\label{section:conclusion}

We present a statistical framework for measuring precise ages of MS stars and
subgiants by combining observables that relate, via different evolutionary
processes, to stellar age.
Specifically, we combined HRD/CMD placement with rotation periods, in a
hierarchical Bayesian model, to age-date stars based on both their hydrogen
burning and magnetic braking history.
The two methods of isochrone fitting and gyrochronology were combined by
taking the product of two likelihoods: one that contains an isochronal model
and the other a gyrochronal one.
We used the MIST stellar evolution models and computed isochronal ages and
likelihoods using the {\tt isochrones} {\it Python} package.
We fit a new broken power law gyrochronology model to the Praesepe cluster
and included a modification recommended by \citet{vansaders2016} that accounts
for weakened magnetic braking at Rossby numbers larger than 2.
The rotation periods of hot stars, cool stars, evolved stars, young stars and
metal poor and rich stars were modeled with a broad log-normal distribution.
We tested \sd\ on simulated data and cluster stars and demonstrated that
combining gyrochronology with isochrone fitting improves the precision of age
estimates for FGK dwarfs by a factor of 3 over isochrone fitting alone,
assuming 5\% measurement uncertainties on rotation periods.
Incorporating rotation periods into stellar evolution models also improves the
precision of the equivalent evolutionary phase (EEP) parameter and, since EEP,
combined with age and metallicity, determines the mass, radius and \logg\ of a
star, this means rotation periods can improve the precision of {\it all}
stellar parameters.
Although V-band extinction is marginalized over during inference, correcting
photometry for dust-extinction before analysis, or including it as a
prior can improve the accuracy of stellar ages measured with \sd.
We also tested \sd\ on a set of 21 \kepler\ asteroseismic stars
\citep{vansaders2016}.
We found that discrepancies between ages measured with \sd\ and ages measured
with asteroseismology are likely produced by differences between the MIST and
BaSTI stellar evolution models.
Asteroseismic and cluster stars provide an opportunity for calibration but
given the high-dimensionality of the gyrochronology relations (\ie\ rotation
period depends on age, mass, metallicity, surface gravity, etc), many stars
with precise ages, spanning a range of properties, are still needed to
reliably calibrate them.

In cases where gyrochronology predicts inaccurate stellar ages it is either
because models are not correctly calibrated, because the rotation periods or
rotation period uncertainties are themselves inaccurate, or because of
rotational outliers.
For example, \sd\ may predict inaccurate ages for stars in close binaries
whose interactions influence their rotation period evolution.
Rotational outliers are often seen in clusters \citep[see \eg][]{douglas2016,
rebull2016, douglas2017, rebull2017} and many of these fall above the main
sequence on a CMD, indicating that they are binaries.
In addition, measured rotation periods may not always be accurate and can, in
many cases, be a harmonic of the true rotation period.
For example, a common rotation period measurement failure mode is to measure
half the true rotation period.
The best way to prevent an erroneous or outlying rotation period from
resulting in an erroneous age measurement is to {\it allow} for outlying
rotation periods using a mixture model, a feature that could be built into
\sd\ in the future.

The optimal way to age-date stars is by combining {\it all} their available
age-related observables.
This could ultimately include activity dating via flare rates and
chromospheric activity indices, kinematic dating and chemical dating.
Of all the established age-dating methods, gyrochronology and isochrone
fitting are two of the most complementary.
The two methods are optimal in different parts of the HRD:
gyrochronology works well for FGK dwarfs and isochrone fitting works well for
subgiants and hot stars, so combining the two methods results in consistently
precise ages across a range of masses, ages and evolutionary stages.
In addition, using both methods at once circumvents the need to decide which
method to use {\it a priori}.
It eliminates the circular process of classifying a star based on its CMD
position (M dwarf, subgiant, etc), then deciding which age-dating method to
use, then inferring an age which itself depends on the classification that was
made.
It is important to infer all stellar properties at once since they all depend
on each other.
\sd\ is applicable to a large number of stars: FGKM dwarfs and subgiants with
a rotation period and broad-band photometry.
This already includes tens-of-thousands of \kepler\ and \ktwo\ stars and could
include millions more from \tess, \lsst, \wfirst, \plato, \gaia, and others in
the future.

The code used in this project is available as a {\it Python} package called
\sd, DOI: \url{ https://doi.org/10.5281/zenodo.2712419}.
It is available for download via Github\footnote{git clone
https://github.com/RuthAngus/stardate.git} or through
PyPI\footnote{pip install stardate\_code}.
Documentation is available at \url{https://stardate.readthedocs.io/en/latest/}.
All code used to produce the figures in this paper is available at
\url{https://github.com/RuthAngus/stardate}.
This paper is based on code with the following Git hash:
f739562c1546e117e9bb217e1732c62b41be8061.

Some of the data presented in this paper were obtained from the Mikulski
Archive for Space Telescopes (MAST).
STScI is operated by the Association of Universities for Research in
Astronomy, Inc., under NASA contract NAS5-26555.
Support for MAST for non-HST data is provided by the NASA Office of Space
Science via grant NNX09AF08G and by other grants and contracts.
This paper includes data collected by the Kepler mission. Funding for the
\Kepler\ mission is provided by the NASA Science Mission directorate.

This work made use of the gaia-kepler.fun crossmatch database created by Megan
Bedell

This work has made use of data from the European Space Agency (ESA) mission
{\it Gaia} (\url{https://www.cosmos.esa.int/gaia}), processed by the {\it Gaia}
Data Processing and Analysis Consortium (DPAC,
\url{https://www.cosmos.esa.int/web/gaia/dpac/consortium}). Funding for the DPAC
has been provided by national institutions, in particular the institutions
participating in the {\it Gaia} Multilateral Agreement.

This research was supported in part by the National Science Foundation under
Grant No. NSF PHY-1748958.


\section{Appendix}
\label{section:appendix}

\subsection*{Priors}
\label{section:priors}

We used the default priors in the {\tt isochrones.py} {\it python} package.
The prior over age was,
\begin{equation}
p(A) = \frac{\log(10) 10^{A}}{10^{10.5} - 10^8}, ~~~ 8 < A < 10.5.
\end{equation}
where $A$, is $\log_{10}(\mathrm{Age~[yrs]})$.
The prior over EEP was uniform with an upper limit of 800.
We found that adding this upper limit reduced some multi-modality caused by
the giant branch and resulted in better performance.
The prior over true bulk metallicity was based on the galactic metallicity
distribution, as inferred using data from the Sloan Digital Sky Survey
\citep{casagrande2011}.
It is the product of a Gaussian that describes the metallicity distribution
over halo stars and two Gaussians that describe the metallicity distribution
in the thin and thick disks:
\begin{eqnarray}
    p(F) =
    & H_F \frac{1}{\sqrt{2\pi\sigma_{\mathrm{halo}}^2}}
    \exp\left(-\frac{(F-\mu_{\mathrm{halo}})^2}{2\sigma_{\mathrm{halo}}}\right)
    \\ \nonumber
    & \times (1-H_F)
    \frac{1}{\xi}
    \left[\frac{0.8}{0.15}\exp\left(-\frac{(F-0.016)^2}{2\times 0.15^2}\right)
    + \frac{0.2}{0.22}\exp\left(-\frac{(F-0.15)^2}{2\times
    0.22^2}\right)\right],
\end{eqnarray}
where $H_F = 0.001$ is the halo fraction, $\mu_\mathrm{halo}$ and
$\sigma_{\mathrm{halo}}$ are the mean and standard deviation of a Gaussian
that describes a probability distribution over metallicity in the halo, and
take values -1.5 and 0.4 respectively.
The two Gaussians inside the square brackets describe probability
distributions over metallicity in the thin and thick disks.
The values of the means and standard deviations in these Gaussians are from
\citet{casagrande2011}.
$\xi$ is the integral of everything in the square brackets from $-\infty$ to
$\infty$ and takes the value $\sim 2.507$.
The prior over distance was,
\begin{equation}
    p(D) = \frac{3}{3000^3} D^2, ~~~ 0 < D < 3000,
\end{equation}
with D in kiloparsecs, and, finally, the prior over extinction was uniform
between zero and one,
\begin{equation}
    p(A_V) = U(0, 1).
\end{equation}

\bibliographystyle{plainnat}
\bibliography{hz.bib}

\begin{thebibliography}{59}
\providecommand{\natexlab}[1]{#1}
\providecommand{\url}[1]{\texttt{#1}}
\expandafter\ifx\csname urlstyle\endcsname\relax
  \providecommand{\doi}[1]{doi: #1}\else
  \providecommand{\doi}{doi: \begingroup \urlstyle{rm}\Url}\fi

\bibitem[{Ag{\"u}eros} et~al.(2018){Ag{\"u}eros}, {Bowsher}, {Bochanski},
  {Cargile}, {Covey}, {Douglas}, {Kraus}, {Kundert}, {Law}, {Ahmadi}, and
  {Arce}]{agueros2018}
M.~A. {Ag{\"u}eros}, E.~C. {Bowsher}, J.~J. {Bochanski}, P.~A. {Cargile}, K.~R.
  {Covey}, S.~T. {Douglas}, A.~{Kraus}, A.~{Kundert}, N.~M. {Law}, A.~{Ahmadi},
  and H.~G. {Arce}.
\newblock {A New Look at an Old Cluster: The Membership, Rotation, and Magnetic
  Activity of Low-mass Stars in the 1.3 Gyr Old Open Cluster NGC 752}.
\newblock \emph{\apj}, 862\penalty0 (1):\penalty0 33, Jul 2018.
\newblock \doi{10.3847/1538-4357/aac6ed}.

\bibitem[{Ag{\"u}eros}(2018)]{agueros2018b}
Marcel {Ag{\"u}eros}.
\newblock {The evolution of cool dwarf spin rates: Data, models, and tensions}.
\newblock In \emph{Cambridge Workshop on Cool Stars, Stellar Systems, and the
  Sun}, Cambridge Workshop on Cool Stars, Stellar Systems, and the Sun,
  page~17, Jul 2018.
\newblock \doi{10.5281/zenodo.1343628}.

\bibitem[{Aigrain} et~al.(2015){Aigrain}, {Llama}, {Ceillier}, {Chagas},
  {Davenport}, {Garc{\'\i}a}, {Hay}, {Lanza}, {McQuillan}, {Mazeh}, {de
  Medeiros}, {Nielsen}, and {Reinhold}]{aigrain2015}
S.~{Aigrain}, J.~{Llama}, T.~{Ceillier}, M.~L.~das {Chagas}, J.~R.~A.
  {Davenport}, R.~A. {Garc{\'\i}a}, K.~L. {Hay}, A.~F. {Lanza}, A.~{McQuillan},
  T.~{Mazeh}, J.~R. {de Medeiros}, M.~B. {Nielsen}, and T.~{Reinhold}.
\newblock {Testing the recovery of stellar rotation signals from Kepler light
  curves using a blind hare-and-hounds exercise}.
\newblock \emph{\mnras}, 450\penalty0 (3):\penalty0 3211--3226, Jul 2015.
\newblock \doi{10.1093/mnras/stv853}.

\bibitem[{Angus} et~al.(2015){Angus}, {Aigrain}, {Foreman-Mackey}, and
  {McQuillan}]{angus2015}
R.~{Angus}, S.~{Aigrain}, D.~{Foreman-Mackey}, and A.~{McQuillan}.
\newblock {Calibrating gyrochronology using Kepler asteroseismic targets}.
\newblock \emph{\mnras}, 450:\penalty0 1787--1798, June 2015.
\newblock \doi{10.1093/mnras/stv423}.

\bibitem[{Angus} et~al.(2018){Angus}, {Morton}, {Aigrain}, {Foreman-Mackey},
  and {Rajpaul}]{angus2018}
Ruth {Angus}, Timothy {Morton}, Suzanne {Aigrain}, Daniel {Foreman-Mackey}, and
  Vinesh {Rajpaul}.
\newblock {Inferring probabilistic stellar rotation periods using Gaussian
  processes}.
\newblock \emph{\mnras}, 474\penalty0 (2):\penalty0 2094--2108, Feb 2018.
\newblock \doi{10.1093/mnras/stx2109}.

\bibitem[{Barnes}(2003)]{barnes2003}
S.~A. {Barnes}.
\newblock {On the Rotational Evolution of Solar- and Late-Type Stars, Its
  Magnetic Origins, and the Possibility of Stellar Gyrochronology}.
\newblock \emph{\apj}, 586:\penalty0 464--479, March 2003.
\newblock \doi{10.1086/367639}.

\bibitem[{Barnes}(2007)]{barnes2007}
S.~A. {Barnes}.
\newblock {Ages for Illustrative Field Stars Using Gyrochronology: Viability,
  Limitations, and Errors}.
\newblock \emph{\apj}, 669:\penalty0 1167--1189, November 2007.
\newblock \doi{10.1086/519295}.

\bibitem[{Barnes}(2010)]{barnes2010}
S.~A. {Barnes}.
\newblock {A Simple Nonlinear Model for the Rotation of Main-sequence Cool
  Stars. I. Introduction, Implications for Gyrochronology, and Color-Period
  Diagrams}.
\newblock \emph{\apj}, 722:\penalty0 222--234, October 2010.
\newblock \doi{10.1088/0004-637X/722/1/222}.

\bibitem[{Brown} et~al.(2011){Brown}, {Latham}, {Everett}, and
  {Esquerdo}]{brown2011}
T.~M. {Brown}, D.~W. {Latham}, M.~E. {Everett}, and G.~A. {Esquerdo}.
\newblock {Kepler Input Catalog: Photometric Calibration and Stellar
  Classification}.
\newblock \emph{\aj}, 142:\penalty0 112, October 2011.
\newblock \doi{10.1088/0004-6256/142/4/112}.

\bibitem[{Casagrande} et~al.(2011){Casagrande}, {Sch{\"o}nrich}, {Asplund},
  {Cassisi}, {Ram{\'\i}rez}, {Mel{\'e}ndez}, {Bensby}, and
  {Feltzing}]{casagrande2011}
L.~{Casagrande}, R.~{Sch{\"o}nrich}, M.~{Asplund}, S.~{Cassisi},
  I.~{Ram{\'\i}rez}, J.~{Mel{\'e}ndez}, T.~{Bensby}, and S.~{Feltzing}.
\newblock {New constraints on the chemical evolution of the solar neighbourhood
  and Galactic disc(s). Improved astrophysical parameters for the
  Geneva-Copenhagen Survey}.
\newblock \emph{\aap}, 530:\penalty0 A138, Jun 2011.
\newblock \doi{10.1051/0004-6361/201016276}.

\bibitem[{Ceillier} et~al.(2016){Ceillier}, {van Saders}, {Garc{\'\i}a},
  {Metcalfe}, {Creevey}, {Mathis}, {Mathur}, {Pinsonneault}, {Salabert}, and
  {Tayar}]{ceillier2016}
T.~{Ceillier}, J.~{van Saders}, R.~A. {Garc{\'\i}a}, T.~S. {Metcalfe},
  O.~{Creevey}, S.~{Mathis}, S.~{Mathur}, M.~H. {Pinsonneault}, D.~{Salabert},
  and J.~{Tayar}.
\newblock {Rotation periods and seismic ages of KOIs - comparison with stars
  without detected planets from Kepler observations}.
\newblock \emph{\mnras}, 456\penalty0 (1):\penalty0 119--125, Feb 2016.
\newblock \doi{10.1093/mnras/stv2622}.

\bibitem[{Choi} et~al.(2016){Choi}, {Dotter}, {Conroy}, {Cantiello}, {Paxton},
  and {Johnson}]{choi2016}
J.~{Choi}, A.~{Dotter}, C.~{Conroy}, M.~{Cantiello}, B.~{Paxton}, and B.~D.
  {Johnson}.
\newblock {Mesa Isochrones and Stellar Tracks (MIST). I. Solar-scaled Models}.
\newblock \emph{\apj}, 823:\penalty0 102, June 2016.
\newblock \doi{10.3847/0004-637X/823/2/102}.

\bibitem[{Christensen-Dalsgaard}(2008{\natexlab{a}})]{christensen-dalsgaard2008a}
J.~{Christensen-Dalsgaard}.
\newblock {ASTEC--the Aarhus STellar Evolution Code}.
\newblock \emph{\apss}, 316:\penalty0 13--24, August 2008{\natexlab{a}}.
\newblock \doi{10.1007/s10509-007-9675-5}.

\bibitem[{Christensen-Dalsgaard}(2008{\natexlab{b}})]{christensen-dalsgaard2008b}
J.~{Christensen-Dalsgaard}.
\newblock {ADIPLS--the Aarhus adiabatic oscillation package}.
\newblock \emph{\apss}, 316:\penalty0 113--120, August 2008{\natexlab{b}}.
\newblock \doi{10.1007/s10509-007-9689-z}.

\bibitem[{Curtis} and {Ag{\"u}eros}(2018)]{curtis2018}
Jason {Curtis} and Marcel {Ag{\"u}eros}.
\newblock {Problems with and Prospects for K dwarf Gyrochronology: Insights
  from the K2 Survey of Ruprecht 147}.
\newblock In \emph{Cambridge Workshop on Cool Stars, Stellar Systems, and the
  Sun}, Cambridge Workshop on Cool Stars, Stellar Systems, and the Sun,
  page~24, Jul 2018.
\newblock \doi{10.5281/zenodo.1403719}.

\bibitem[{Curtis} et~al.(2019){Curtis}, {Ag{\"u}eros}, {Douglas}, and
  {Meibom}]{curtis2019}
Jason~Lee {Curtis}, Marcel~A. {Ag{\"u}eros}, Stephanie~T. {Douglas}, and
  S{\o}ren {Meibom}.
\newblock {A Temporary Epoch of Stalled Spin-Down for Low-Mass Stars: Insights
  from NGC 6811 with Gaia and Kepler}.
\newblock \emph{arXiv e-prints}, art. arXiv:1905.06869, May 2019.

\bibitem[{Donahue} et~al.(1996){Donahue}, {Saar}, and {Baliunas}]{donahue1996}
Robert~A. {Donahue}, Steven~H. {Saar}, and Sallie~L. {Baliunas}.
\newblock {A Relationship between Mean Rotation Period in Lower Main-Sequence
  Stars and Its Observed Range}.
\newblock \emph{\apj}, 466:\penalty0 384, Jul 1996.
\newblock \doi{10.1086/177517}.

\bibitem[{Dotter}(2016)]{dotter2016}
A.~{Dotter}.
\newblock {MESA Isochrones and Stellar Tracks (MIST) 0: Methods for the
  Construction of Stellar Isochrones}.
\newblock \emph{\apjs}, 222:\penalty0 8, January 2016.
\newblock \doi{10.3847/0067-0049/222/1/8}.

\bibitem[{Dotter} et~al.(2008){Dotter}, {Chaboyer}, {Jevremovi{\'c}}, {Kostov},
  {Baron}, and {Ferguson}]{dotter2008}
A.~{Dotter}, B.~{Chaboyer}, D.~{Jevremovi{\'c}}, V.~{Kostov}, E.~{Baron}, and
  J.~W. {Ferguson}.
\newblock {The Dartmouth Stellar Evolution Database}.
\newblock \emph{\apjs}, 178:\penalty0 89--101, September 2008.
\newblock \doi{10.1086/589654}.

\bibitem[{Douglas} et~al.(2016){Douglas}, {Ag{\"u}eros}, {Covey}, {Cargile},
  {Barclay}, {Cody}, {Howell}, and {Kopytova}]{douglas2016}
S.~T. {Douglas}, M.~A. {Ag{\"u}eros}, K.~R. {Covey}, P.~A. {Cargile},
  T.~{Barclay}, A.~{Cody}, S.~B. {Howell}, and T.~{Kopytova}.
\newblock {K2 Rotation Periods for Low-mass Hyads and the Implications for
  Gyrochronology}.
\newblock \emph{\apj}, 822:\penalty0 47, May 2016.
\newblock \doi{10.3847/0004-637X/822/1/47}.

\bibitem[{Douglas} et~al.(2017){Douglas}, {Ag{\"u}eros}, {Covey}, and
  {Kraus}]{douglas2017}
S.~T. {Douglas}, M.~A. {Ag{\"u}eros}, K.~R. {Covey}, and A.~{Kraus}.
\newblock {Poking the Beehive from Space: K2 Rotation Periods for Praesepe}.
\newblock \emph{\apj}, 842:\penalty0 83, June 2017.
\newblock \doi{10.3847/1538-4357/aa6e52}.

\bibitem[{Epstein} and {Pinsonneault}(2014)]{epstein2014}
C.~R. {Epstein} and M.~H. {Pinsonneault}.
\newblock {How Good a Clock is Rotation? The Stellar Rotation-Mass-Age
  Relationship for Old Field Stars}.
\newblock \emph{\apj}, 780:\penalty0 159, January 2014.
\newblock \doi{10.1088/0004-637X/780/2/159}.

\bibitem[{Evans} et~al.(2017){Evans}, {Riello}, {De Angeli}, {Busso}, {van
  Leeuwen}, {Jordi}, {Fabricius}, {Brown}, {Carrasco}, {Voss}, {Weiler},
  {Montegriffo}, {Cacciari}, {Burgess}, and {Osborne}]{evans2017}
D.~W. {Evans}, M.~{Riello}, F.~{De Angeli}, G.~{Busso}, F.~{van Leeuwen},
  C.~{Jordi}, C.~{Fabricius}, A.~G.~A. {Brown}, J.~M. {Carrasco}, H.~{Voss},
  M.~{Weiler}, P.~{Montegriffo}, C.~{Cacciari}, P.~{Burgess}, and P.~{Osborne}.
\newblock {Gaia Data Release 1. Validation of the photometry}.
\newblock \emph{\aap}, 600:\penalty0 A51, April 2017.
\newblock \doi{10.1051/0004-6361/201629241}.

\bibitem[{Foreman-Mackey} et~al.(2013){Foreman-Mackey}, {Hogg}, {Lang}, and
  {Goodman}]{foreman-mackey2013}
D.~{Foreman-Mackey}, D.~W. {Hogg}, D.~{Lang}, and J.~{Goodman}.
\newblock {emcee: The MCMC Hammer}.
\newblock \emph{\pasp}, 125:\penalty0 306, March 2013.
\newblock \doi{10.1086/670067}.

\bibitem[{Fossati} et~al.(2008){Fossati}, {Bagnulo}, {Landstreet}, {Wade},
  {Kochukhov}, {Monier}, {Weiss}, and {Gebran}]{fossati2008}
L.~{Fossati}, S.~{Bagnulo}, J.~{Landstreet}, G.~{Wade}, O.~{Kochukhov},
  R.~{Monier}, W.~{Weiss}, and M.~{Gebran}.
\newblock {The effect of rotation on the abundances of the chemical elements of
  the A-type stars in the Praesepe cluster}.
\newblock \emph{\aap}, 483:\penalty0 891--902, June 2008.
\newblock \doi{10.1051/0004-6361:200809467}.

\bibitem[{Gaia Collaboration} et~al.(2018){Gaia Collaboration}, {Brown},
  {Vallenari}, {Prusti}, {de Bruijne}, {Babusiaux}, {Bailer-Jones}, {Biermann},
  {Evans}, {Eyer}, and et~al.]{brown2018}
{Gaia Collaboration}, A.~G.~A. {Brown}, A.~{Vallenari}, T.~{Prusti}, J.~H.~J.
  {de Bruijne}, C.~{Babusiaux}, C.~A.~L. {Bailer-Jones}, M.~{Biermann}, D.~W.
  {Evans}, L.~{Eyer}, and et~al.
\newblock {Gaia Data Release 2. Summary of the contents and survey properties}.
\newblock \emph{\aap}, 616:\penalty0 A1, August 2018.
\newblock \doi{10.1051/0004-6361/201833051}.

\bibitem[{Gallet} and {Bouvier}(2015)]{gallet2015}
F.~{Gallet} and J.~{Bouvier}.
\newblock {Improved angular momentum evolution model for solar-like stars. II.
  Exploring the mass dependence}.
\newblock \emph{\aap}, 577:\penalty0 A98, May 2015.
\newblock \doi{10.1051/0004-6361/201525660}.

\bibitem[{Garc{\'{\i}}a} et~al.(2014){Garc{\'{\i}}a}, {Ceillier}, {Salabert},
  {Mathur}, {van Saders}, {Pinsonneault}, {Ballot}, {Beck}, {Bloemen},
  {Campante}, {Davies}, {do Nascimento}, {Mathis}, {Metcalfe}, {Nielsen},
  {Su{\'a}rez}, {Chaplin}, {Jim{\'e}nez}, and {Karoff}]{garcia2014}
R.~A. {Garc{\'{\i}}a}, T.~{Ceillier}, D.~{Salabert}, S.~{Mathur}, J.~L. {van
  Saders}, M.~{Pinsonneault}, J.~{Ballot}, P.~G. {Beck}, S.~{Bloemen}, T.~L.
  {Campante}, G.~R. {Davies}, J.-D. {do Nascimento}, Jr., S.~{Mathis}, T.~S.
  {Metcalfe}, M.~B. {Nielsen}, J.~C. {Su{\'a}rez}, W.~J. {Chaplin},
  A.~{Jim{\'e}nez}, and C.~{Karoff}.
\newblock {Rotation and magnetism of Kepler pulsating solar-like stars. Towards
  asteroseismically calibrated age-rotation relations}.
\newblock \emph{\aap}, 572:\penalty0 A34, December 2014.
\newblock \doi{10.1051/0004-6361/201423888}.

\bibitem[{Hidalgo} et~al.(2018){Hidalgo}, {Pietrinferni}, {Cassisi}, {Salaris},
  {Mucciarelli}, {Savino}, {Aparicio}, {Silva Aguirre}, and
  {Verma}]{hidalgo2018}
Sebastian~L. {Hidalgo}, Adriano {Pietrinferni}, Santi {Cassisi}, Maurizio
  {Salaris}, Alessio {Mucciarelli}, Alessandro {Savino}, Antonio {Aparicio},
  Victor {Silva Aguirre}, and Kuldeep {Verma}.
\newblock {The Updated BaSTI Stellar Evolution Models and Isochrones. I.
  Solar-scaled Calculations}.
\newblock \emph{\apj}, 856\penalty0 (2):\penalty0 125, Apr 2018.
\newblock \doi{10.3847/1538-4357/aab158}.

\bibitem[{Huber} et~al.(2016){Huber}, {Bryson}, {Haas}, {Barclay}, {Barentsen},
  {Howell}, {Sharma}, {Stello}, and {Thompson}]{huber2016}
Daniel {Huber}, Stephen~T. {Bryson}, Michael~R. {Haas}, Thomas {Barclay}, Geert
  {Barentsen}, Steve~B. {Howell}, Sanjib {Sharma}, Dennis {Stello}, and
  Susan~E. {Thompson}.
\newblock {The K2 Ecliptic Plane Input Catalog (EPIC) and Stellar
  Classifications of 138,600 Targets in Campaigns 1-8}.
\newblock \emph{The Astrophysical Journal Supplement Series}, 224:\penalty0 2,
  May 2016.
\newblock \doi{10.3847/0067-0049/224/1/2}.

\bibitem[{Irwin} and {Bouvier}(2009)]{irwin2009}
J.~{Irwin} and J.~{Bouvier}.
\newblock {The rotational evolution of low-mass stars}.
\newblock In E.~E. {Mamajek}, D.~R. {Soderblom}, and R.~F.~G. {Wyse}, editors,
  \emph{The Ages of Stars}, volume 258 of \emph{IAU Symposium}, pages 363--374,
  June 2009.
\newblock \doi{10.1017/S1743921309032025}.

\bibitem[{Kawaler}(1988)]{kawaler1988}
S.~D. {Kawaler}.
\newblock {Angular momentum loss in low-mass stars}.
\newblock \emph{\apj}, 333:\penalty0 236--247, October 1988.
\newblock \doi{10.1086/166740}.

\bibitem[{Kawaler}(1989)]{kawaler1989}
S.~D. {Kawaler}.
\newblock {Rotational dating of middle-aged stars}.
\newblock \emph{\apjl}, 343:\penalty0 L65--L68, August 1989.
\newblock \doi{10.1086/185512}.

\bibitem[{Krishnamurthi} et~al.(1997){Krishnamurthi}, {Pinsonneault}, {Barnes},
  and {Sofia}]{krishnamurthi1997}
A.~{Krishnamurthi}, M.~H. {Pinsonneault}, S.~{Barnes}, and S.~{Sofia}.
\newblock {Theoretical Models of the Angular Momentum Evolution of Solar-Type
  Stars}.
\newblock \emph{\apj}, 480:\penalty0 303--323, May 1997.
\newblock \doi{10.1086/303958}.

\bibitem[{Mamajek} and {Hillenbrand}(2008)]{mamajek2008}
E.~E. {Mamajek} and L.~A. {Hillenbrand}.
\newblock {Improved Age Estimation for Solar-Type Dwarfs Using
  Activity-Rotation Diagnostics}.
\newblock \emph{\apj}, 687:\penalty0 1264--1293, November 2008.
\newblock \doi{10.1086/591785}.

\bibitem[{Mathur} et~al.(2012){Mathur}, {Metcalfe}, {Woitaszek}, {Bruntt},
  {Verner}, {Christensen-Dalsgaard}, {Creevey}, {Do{\v{g}}an}, {Basu}, and
  {Karoff}]{mathur2012}
S.~{Mathur}, T.~S. {Metcalfe}, M.~{Woitaszek}, H.~{Bruntt}, G.~A. {Verner},
  J.~{Christensen-Dalsgaard}, O.~L. {Creevey}, G.~{Do{\v{g}}an}, S.~{Basu}, and
  C.~{Karoff}.
\newblock {A Uniform Asteroseismic Analysis of 22 Solar-type Stars Observed by
  Kepler}.
\newblock \emph{\apj}, 749\penalty0 (2):\penalty0 152, Apr 2012.
\newblock \doi{10.1088/0004-637X/749/2/152}.

\bibitem[{Matt} et~al.(2012){Matt}, {MacGregor}, {Pinsonneault}, and
  {Greene}]{matt2012}
S.~P. {Matt}, K.~B. {MacGregor}, M.~H. {Pinsonneault}, and T.~P. {Greene}.
\newblock {Magnetic Braking Formulation for Sun-like Stars: Dependence on
  Dipole Field Strength and Rotation Rate}.
\newblock \emph{\apjl}, 754:\penalty0 L26, August 2012.
\newblock \doi{10.1088/2041-8205/754/2/L26}.

\bibitem[{Matt} et~al.(2015){Matt}, {Brun}, {Baraffe}, {Bouvier}, and
  {Chabrier}]{matt2015}
S.~P. {Matt}, A.~S. {Brun}, I.~{Baraffe}, J.~{Bouvier}, and G.~{Chabrier}.
\newblock {The Mass-dependence of Angular Momentum Evolution in Sun-like
  Stars}.
\newblock \emph{\apjl}, 799:\penalty0 L23, January 2015.
\newblock \doi{10.1088/2041-8205/799/2/L23}.

\bibitem[{McQuillan} et~al.(2014){McQuillan}, {Mazeh}, and
  {Aigrain}]{mcquillan2014}
A.~{McQuillan}, T.~{Mazeh}, and S.~{Aigrain}.
\newblock {Rotation Periods of 34,030 Kepler Main-sequence Stars: The Full
  Autocorrelation Sample}.
\newblock \emph{\apjs}, 211:\penalty0 24, April 2014.
\newblock \doi{10.1088/0067-0049/211/2/24}.

\bibitem[{Meibom} et~al.(2015){Meibom}, {Barnes}, {Platais}, {Gilliland},
  {Latham}, and {Mathieu}]{meibom2015}
S.~{Meibom}, S.~A. {Barnes}, I.~{Platais}, R.~L. {Gilliland}, D.~W. {Latham},
  and R.~D. {Mathieu}.
\newblock {A spin-down clock for cool stars from observations of a
  2.5-billion-year-old cluster}.
\newblock \emph{\nat}, 517:\penalty0 589--591, January 2015.
\newblock \doi{10.1038/nature14118}.

\bibitem[{Metcalfe} et~al.(2009){Metcalfe}, {Creevey}, and
  {Christensen-Dalsgaard}]{metcalfe2009}
T.~S. {Metcalfe}, O.~L. {Creevey}, and J.~{Christensen-Dalsgaard}.
\newblock {A Stellar Model-fitting Pipeline for Asteroseismic Data from the
  Kepler Mission}.
\newblock \emph{\apj}, 699:\penalty0 373--382, July 2009.
\newblock \doi{10.1088/0004-637X/699/1/373}.

\bibitem[{Metcalfe} et~al.(2012){Metcalfe}, {Mathur}, {Do{\u{g}}an}, and
  {Woitaszek}]{metcalfe2012}
T.~S. {Metcalfe}, S.~{Mathur}, G.~{Do{\u{g}}an}, and M.~{Woitaszek}.
\newblock {First Results from the Asteroseismic Modeling Portal}.
\newblock In H.~{Shibahashi}, M.~{Takata}, and A.~E. {Lynas-Gray}, editors,
  \emph{Progress in Solar/Stellar Physics with Helio- and Asteroseismology},
  volume 462 of \emph{Astronomical Society of the Pacific Conference Series},
  page 213, Sep 2012.

\bibitem[{Metcalfe} et~al.(2014){Metcalfe}, {Creevey}, {Do{\u{g}}an}, {Mathur},
  {Xu}, {Bedding}, {Chaplin}, {Christensen-Dalsgaard}, {Karoff}, and
  {Trampedach}]{metcalfe2014}
T.~S. {Metcalfe}, O.~L. {Creevey}, G.~{Do{\u{g}}an}, S.~{Mathur}, H.~{Xu},
  T.~R. {Bedding}, W.~J. {Chaplin}, J.~{Christensen-Dalsgaard}, C.~{Karoff},
  and R.~{Trampedach}.
\newblock {Properties of 42 Solar-type Kepler Targets from the Asteroseismic
  Modeling Portal}.
\newblock \emph{\apjs}, 214\penalty0 (2):\penalty0 27, Oct 2014.
\newblock \doi{10.1088/0067-0049/214/2/27}.

\bibitem[{Morton}(2015)]{isochrones}
T.~D. {Morton}.
\newblock {isochrones: Stellar model grid package}.
\newblock Astrophysics Source Code Library, March 2015.

\bibitem[{Pietrinferni} et~al.(2004){Pietrinferni}, {Cassisi}, {Salaris}, and
  {Castelli}]{pietrinferni2004}
Adriano {Pietrinferni}, Santi {Cassisi}, Maurizio {Salaris}, and Fiorella
  {Castelli}.
\newblock {A Large Stellar Evolution Database for Population Synthesis Studies.
  I. Scaled Solar Models and Isochrones}.
\newblock \emph{\apj}, 612\penalty0 (1):\penalty0 168--190, Sep 2004.
\newblock \doi{10.1086/422498}.

\bibitem[{Pinsonneault} et~al.(1989){Pinsonneault}, {Kawaler}, {Sofia}, and
  {Demarque}]{pinsonneault1989}
M.~H. {Pinsonneault}, S.~D. {Kawaler}, S.~{Sofia}, and P.~{Demarque}.
\newblock {Evolutionary models of the rotating sun}.
\newblock \emph{\apj}, 338:\penalty0 424--452, March 1989.
\newblock \doi{10.1086/167210}.

\bibitem[{Pont} and {Eyer}(2004)]{pont2004}
Fr{\'e}d{\'e}ric {Pont} and Laurent {Eyer}.
\newblock {Isochrone ages for field dwarfs: method and application to the
  age-metallicity relation}.
\newblock \emph{\mnras}, 351\penalty0 (2):\penalty0 487--504, Jun 2004.
\newblock \doi{10.1111/j.1365-2966.2004.07780.x}.

\bibitem[{Rebull} et~al.(2016){Rebull}, {Stauffer}, {Bouvier}, {Cody},
  {Hillenbrand}, {Soderblom}, {Valenti}, {Barrado}, {Bouy}, {Ciardi},
  {Pinsonneault}, {Stassun}, {Micela}, {Aigrain}, {Vrba}, {Somers},
  {Christiansen}, {Gillen}, and {Collier Cameron}]{rebull2016}
L.~M. {Rebull}, J.~R. {Stauffer}, J.~{Bouvier}, A.~M. {Cody}, L.~A.
  {Hillenbrand}, D.~R. {Soderblom}, J.~{Valenti}, D.~{Barrado}, H.~{Bouy},
  D.~{Ciardi}, M.~{Pinsonneault}, K.~{Stassun}, G.~{Micela}, S.~{Aigrain},
  F.~{Vrba}, G.~{Somers}, J.~{Christiansen}, E.~{Gillen}, and A.~{Collier
  Cameron}.
\newblock {Rotation in the Pleiades with K2. I. Data and First Results}.
\newblock \emph{\aj}, 152:\penalty0 113, November 2016.
\newblock \doi{10.3847/0004-6256/152/5/113}.

\bibitem[{Rebull} et~al.(2017){Rebull}, {Stauffer}, {Hillenbrand}, {Cody},
  {Bouvier}, {Soderblom}, {Pinsonneault}, and {Hebb}]{rebull2017}
L.~M. {Rebull}, J.~R. {Stauffer}, L.~A. {Hillenbrand}, A.~M. {Cody},
  J.~{Bouvier}, D.~R. {Soderblom}, M.~{Pinsonneault}, and L.~{Hebb}.
\newblock {Rotation of Late-type Stars in Praesepe with K2}.
\newblock \emph{\apj}, 839:\penalty0 92, April 2017.
\newblock \doi{10.3847/1538-4357/aa6aa4}.

\bibitem[{Silva Aguirre} et~al.(2015){Silva Aguirre}, {Davies}, {Basu},
  {Christensen-Dalsgaard}, {Creevey}, {Metcalfe}, {Bedding}, {Casagrande},
  {Handberg}, {Lund}, {Nissen}, {Chaplin}, {Huber}, {Serenelli}, {Stello}, {Van
  Eylen}, {Campante}, {Elsworth}, {Gilliland}, {Hekker}, {Karoff}, {Kawaler},
  {Kjeldsen}, and {Lundkvist}]{silvaaguirre2015}
V.~{Silva Aguirre}, G.~R. {Davies}, S.~{Basu}, J.~{Christensen-Dalsgaard},
  O.~{Creevey}, T.~S. {Metcalfe}, T.~R. {Bedding}, L.~{Casagrande},
  R.~{Handberg}, M.~N. {Lund}, P.~E. {Nissen}, W.~J. {Chaplin}, D.~{Huber},
  A.~M. {Serenelli}, D.~{Stello}, V.~{Van Eylen}, T.~L. {Campante},
  Y.~{Elsworth}, R.~L. {Gilliland}, S.~{Hekker}, C.~{Karoff}, S.~D. {Kawaler},
  H.~{Kjeldsen}, and M.~S. {Lundkvist}.
\newblock {Ages and fundamental properties of Kepler exoplanet host stars from
  asteroseismology}.
\newblock \emph{\mnras}, 452:\penalty0 2127--2148, September 2015.
\newblock \doi{10.1093/mnras/stv1388}.

\bibitem[{Skumanich}(1972)]{skumanich1972}
A.~{Skumanich}.
\newblock {Time Scales for CA II Emission Decay, Rotational Braking, and
  Lithium Depletion}.
\newblock \emph{\apj}, 171:\penalty0 565, February 1972.
\newblock \doi{10.1086/151310}.

\bibitem[{Soderblom}(2010)]{soderblom2010}
D.~R. {Soderblom}.
\newblock {The Ages of Stars}.
\newblock \emph{\araa}, 48:\penalty0 581--629, September 2010.
\newblock \doi{10.1146/annurev-astro-081309-130806}.

\bibitem[{Somers} et~al.(2017){Somers}, {Stauffer}, {Rebull}, {Cody}, and
  {Pinsonneault}]{somers2017}
Garrett {Somers}, John {Stauffer}, Luisa {Rebull}, Ann~Marie {Cody}, and Marc
  {Pinsonneault}.
\newblock {M Dwarf Rotation from the K2 Young Clusters to the Field. I. A
  Mass-Rotation Correlation at 10 Myr}.
\newblock \emph{\apj}, 850\penalty0 (2):\penalty0 134, Dec 2017.
\newblock \doi{10.3847/1538-4357/aa93ed}.

\bibitem[{Tayar} and {Pinsonneault}(2018)]{tayar2018}
Jamie {Tayar} and Marc~H. {Pinsonneault}.
\newblock {Testing Angular Momentum Transport and Wind Loss in
  Intermediate-mass Core-helium Burning Stars}.
\newblock \emph{\apj}, 868\penalty0 (2):\penalty0 150, Dec 2018.
\newblock \doi{10.3847/1538-4357/aae979}.

\bibitem[{van Saders} and {Pinsonneault}(2013)]{vansaders2013}
J.~L. {van Saders} and M.~H. {Pinsonneault}.
\newblock {Fast Star, Slow Star; Old Star, Young Star: Subgiant Rotation as a
  Population and Stellar Physics Diagnostic}.
\newblock \emph{\apj}, 776:\penalty0 67, October 2013.
\newblock \doi{10.1088/0004-637X/776/2/67}.

\bibitem[{van Saders} et~al.(2016){van Saders}, {Ceillier}, {Metcalfe}, {Silva
  Aguirre}, {Pinsonneault}, {Garc{\'{\i}}a}, {Mathur}, and
  {Davies}]{vansaders2016}
J.~L. {van Saders}, T.~{Ceillier}, T.~S. {Metcalfe}, V.~{Silva Aguirre}, M.~H.
  {Pinsonneault}, R.~A. {Garc{\'{\i}}a}, S.~{Mathur}, and G.~R. {Davies}.
\newblock {Weakened magnetic braking as the origin of anomalously rapid
  rotation in old field stars}.
\newblock \emph{\nat}, 529:\penalty0 181--184, January 2016.
\newblock \doi{10.1038/nature16168}.

\bibitem[{van Saders} et~al.(2018){van Saders}, {Pinsonneault}, and
  {Barbieri}]{vansaders2018}
J.~L. {van Saders}, M.~H. {Pinsonneault}, and M.~{Barbieri}.
\newblock {Forward Modeling of the Kepler Stellar Rotation Period Distribution:
  Interpreting Periods from Mixed and Biased Stellar Populations}.
\newblock \emph{ArXiv e-prints}, March 2018.

\bibitem[{Wright} et~al.(2011){Wright}, {Drake}, {Mamajek}, and
  {Henry}]{wright2011}
N.~J. {Wright}, J.~J. {Drake}, E.~E. {Mamajek}, and G.~W. {Henry}.
\newblock {The Stellar-activity-Rotation Relationship and the Evolution of
  Stellar Dynamos}.
\newblock \emph{\apj}, 743:\penalty0 48, December 2011.
\newblock \doi{10.1088/0004-637X/743/1/48}.

\bibitem[{Yi} et~al.(2001){Yi}, {Demarque}, {Kim}, {Lee}, {Ree}, {Lejeune}, and
  {Barnes}]{yi2001}
S.~{Yi}, P.~{Demarque}, Y.-C. {Kim}, Y.-W. {Lee}, C.~H. {Ree}, T.~{Lejeune},
  and S.~{Barnes}.
\newblock {Toward Better Age Estimates for Stellar Populations: The Y$^{2}$
  Isochrones for Solar Mixture}.
\newblock \emph{\apjs}, 136:\penalty0 417--437, October 2001.
\newblock \doi{10.1086/321795}.

\end{thebibliography}
\end{document}